\documentclass[a4paper]{article}
\pdfoutput=1

\usepackage{amsmath,amsfonts,amssymb}
\usepackage{amsthm}
\usepackage{enumerate}
\usepackage{graphicx}
\usepackage{rotating}
\usepackage{hyperref}
\usepackage{color}

\def\ds{\displaystyle}
\def\bea{\begin{array}{c}}
\def\ea{\end{array}}
\def\be{\begin{equation}\bea\ds}
\def\ee{\ea\end{equation}}
\def\bee{\begin{equation}\begin{array}{rcl}\ds}
\def\eee{\end{array}\end{equation}}

\def\tr{{\rm Tr}\,}

\def\arccot{{\rm arccot}\,}
\def\arcsinh{{\rm arcsinh}\,}

\def\d{{\rm d}}

\title{\bf Studies of Boundary Entropy in AdS/BCFT}

\author{Arthur G.~Cavalcanti$^{a}$, Dmitry Melnikov$^{b,c}$ and Madson R.~O.~Silva$^{b,d}$}

\begin{document}

\maketitle

\vspace{-5cm}
\hfill{ITEP-TH-21/18}
\vspace{5cm}

\vspace{6pt}
\begin{center}
\textit{\small $^a$  Department of Theoretical and Experimental Physics, Federal University of Rio Grande do Norte, \\ Campus Universit\'ario, Lagoa Nova, Natal-RN  59078-970, Brazil}
\\ \vspace{6pt}
\textit{\small $^b$  International Institute of Physics, Federal University of Rio Grande do Norte, \\ Campus Universit\'ario, Lagoa Nova, Natal-RN  59078-970, Brazil}\\ \vspace{6pt}
\textit{\small $^c$  Institute for Theoretical and Experimental Physics, \\B.~Cheremushkinskaya 25, Moscow 117218, Russia}
\\ \vspace{6pt}
\textit{\small $^d$  Federal University of Maranh\~ao, Campus Balsas,  \\ rua Jos\'e Le\~ao 484, Balsas-MA 65800-000, Brazil}
\\ \vspace{2cm}
\end{center}

\begin{abstract}
In this paper we review the AdS/BCFT proposal of T.~Takayanagi for holographic description of systems with boundaries, in particular, boundary conformal field theories (BCFTs). Motivated by better understanding of the proposed duality we employ entanglement entropy as a probe of familiar properties of impurities and defects. Using the dual gravity description, we check that in two spacetime dimensions the impurity entropy does not depend on a particular state of the theory, which is a well-known CFT result. In three dimensions different, and not necessarily equivalent, definitions of the defect entropy can be given. We compute the entanglement entropy of a line defect at finite temperature and compare it with earlier calculations of the thermodynamical entropy. The results indicate that the entanglement entropy flows to the definition of the entropy as the Bekenstein-Hawking entropy associated to a portion of the black horizon, which we call impurity ``shadow". Geometric configurations, which we discuss, provide examples of RG flows of the defect entropies. We outline the connection between the geometric picture of the RG flows and examples of lattice calculations. We also discuss some new generalizations of the AdS/BCFT geometries.
\end{abstract}

\newpage

\section{Introduction}

Boundary conditions often appear as a mathematical nuisance in our study of physical systems. Meanwhile, physics often depends on their choice in a crucial way. Proper choice of boundary conditions is an important step in setting up a correct theory of a physical phenomenon. In extreme cases boundaries completely encode the phenomenon. It is customary to refer to such cases as to the bulk-to-boundary, or holographic, correspondence. The prime examples, in which bulk-to-boundary correspondence is a pronounced feature, are topological states of matter and AdS/CFT correspondence.

Conformal field theories (CFTs) in $1+1$ dimensions provide a good starting point for investigating the role of boundary conditions in bulk physics and the bulk-to-boundary correspondence. On one hand, different boundary conditions can be understood as perturbations of CFTs by different operator insertions at the boundaries,
\be
\label{CFTdeform}
\delta S_{\rm CFT} \ = \ \sum\limits_i\int d^2x\ g_i\,{\cal O}_i(x_i)\,.
\ee
Working in $1+1$ dimensions here has the advantage of powerful CFT and integrability methods being available, which implies a variety of  analytical results.

On the other hand, $1+1$-dimensional models have been a subject of much of recent efforts in understanding quantum gravity through lower-dimensional versions of AdS/CFT correspondence~\cite{AdS/CFT}. The latter viewpoint gives another perspective on the idea of bulk-to-boundary correspondence, in which the $1+1$-dimensional system itself can be viewed as encoding an emerging quantum gravity theory in a higher dimension. In this work we will consider a set of models where both perspectives are present.

In a 2011 paper, Tadashi Takayanagi put forward a proposal of a dual gravity construction for $d$-dimensional theories with boundaries~\cite{Takayanagi:2011zk}. Dubbed \emph{AdS/BCFT correspondence}, where $B$ stays for boundary, it was aimed, in particular, at a class of boundary conditions partially preserving conformal symmetry. Such systems were extensively studied in the literature by conventional quantum field theory and CFT methods~\cite{Cardy:2004hm}. Therefore, one should in general expect that the results of those studies can be compared to the predictions of the AdS/BCFT model.

The expectation looks correct so far. One of the probes of the proposed constructions, used in~\cite{Takayanagi:2011zk} and in subsequent papers (see~\cite{AdS/BCFT2,HoloKondo,Chu:2017aab,AdSBCFTexamples} and references therein), is the entropy associated with the boundary degrees of freedom (boundary entropy). In $1+1$ dimensions this entropy does not depend on temperature and can be computed either from equilibrium thermodynamics, or from the entanglement entropy of an interval, containing the impurity. The entropy is usually expressed in terms of a \emph{$g$-function}
\be
\label{gFunction}
S_{\rm bry} \ = \ \sum\limits_i^{\rm all~ boundaries}\log g_i\,,
\ee
which provides a measure of the number of boundary degrees of freedom at each boundary $i$. In analogy with the central charge of $2D$ CFTs, $g$-function is expected to be a non-decreasing function of the energy scale, which is the statement of the so-called g-theorem~\cite{Freedan}.

In the $1+1$-dimensional AdS/BCFT model different calculations do show agreement with the CFT expectations. Boundary entropy is well-defined, independent from temperature and the size of the bulk system, exactly as it could be expected from CFT, \emph{cf.}~\cite{Universality}. Different methods of computation lead to the same result for the holographic $g$-function, which we cast in the form
\be
\label{BEntropy}
S_{\rm bry} \ = \ \frac{c}{6}\,\log\cot\frac{\theta}{2}\,.
\ee
Here $\theta$ is a parameter characterizing the boundary condition -- the BCFT state $|\theta\rangle$. In the geometric picture $\theta$ is the local angle, at which the CFT boundary is extended inside the AdS bulk (see figure~\ref{fig:NMQP}). Consequently, one can construct a geometric proof of the g-theorem, which relies on some physically motivated restrictions on the bulk configuration known as (null or weak) energy conditions~\cite{Takayanagi:2011zk}.

Note that equation~(\ref{BEntropy}) represents a CFT result for the boundary entropy. In other words, for a generic system, it should correspond to a fixed-point value, to which the thermodynamic, or entanglement entropy may renormalize at low energies. This RG flow has a rather simple geometric interpretation. In the case of entanglement entropy, computed holographically via the Ryu and Takayanagi (RT) formula~\cite{RT}, it reduces to a statement about the growth of a segment of the geometric distance (more generally area) associated with the boundary state $|\theta\rangle$, with the growth of the size of the entangled subsystem: this geometric quantity should decrease with increase of the size, which can be seen as an alternative way to state the g-theorem.

We consider several examples of geometric configurations, which are either at fixed point, or exhibit an RG flow in $1+1$ and $2+1$ dimensions. First, we observe that, in lower dimension the Takayanagi's boundary condition automatically ensures that entropy~(\ref{BEntropy}) is independent from the details of the geometric background. Technically this happens because the extensions of the defect to the bulk, determined by the boundary conditions, are always orthogonal to the geodesic lines anchored on the boundary. This property does not hold in higher dimensions. Consequently, the simplest example in $2+1$-dimensions shows an RG flow.

In $2+1$ dimensions we are interested in comparing the entropy of the defect computed by the entanglement entropy with the thermodynamical entropy associated with the defect. The thermodynamical derivation of the entropy of a line defect in $2+1$ dimensions was performed by some of us in~\cite{Magan:2014dwa}. That work proposed two definitions of the entropy: one based on the explicit calculation of the thermodynamical potentials in the gravity system, and another one, derived from a fluid (thermo) dynamics on the defect brane. In particular, the second definition is consistent with entropy being a Bekenstein-Hawking factor multiplying the area of the defect brane projected onto horizon (impurity shadow). Our study indicates that at finite temperature, the entanglement entropy flows to that second definition.

We note that the Takayanagi's boundary conditions (the associated geometrical parameter $\theta$) and the boundary conditions and parameters on the BCFT side, have not so far been related explicitly. In connection to this problem we discuss a known property of RG flows associated with impurity deformations: a relevant perturbation, inserted locally in a given sample, may effectively split it into two unentangled pieces. (This was first proposed in~\cite{BulkImp} and later confirmed by a DMRG calculations in spin chains, \emph{e.g.}~\cite{Affleck:review}.) The fixed points of the RG flow induced by relevant/irrelevant operators can be illustrated by simple geometric pictures, distinguished by values of $\theta$. For example, $\theta>\pi/2$ corresponds to a relevant perturbation: the region of the bulk encoding the impurity (impurity wedge) falls in the interior of the so-called entanglement wedge so that the RT entanglement entropy is zero.

We conclude that geometric parameter $\theta$ can be translated to the coupling constant $g$ that sources deformation~(\ref{CFTdeform}) inserted at the boundary of the dual $1+1$-dimensional CFT. In terms of the BCFT, it is a source of the operator ${\cal O}_\theta$, which creates the boundary state, ${\cal O}_\theta|0\rangle = |\theta\rangle$. However, we leave a more precise definition of  boundary deformation operator ${\cal O}_\theta$ and boundary state $|\theta\rangle$ for a future work. This paper is organized as follows.

In section~\ref{sec:AdSBCFT} we review the AdS/BCFT proposal introduced by Takayanagi in~\cite{Takayanagi:2011zk}. After describing the geometric construction we review the most relevant examples. Other examples, including some new solutions, are relegated to appendix section~\ref{appendix}. We mainly work with the Poincar\'e patch of AdS space, but we review the global patch version in appendix~\ref{global}.

In section~\ref{sec:physics} we discuss physical properties of impurities following from the AdS/BCFT construction of section~\ref{sec:AdSBCFT}. First, we remind the formulas for the thermodynamic entropies associated with impurities and defects in different dimensions. For $1+1$-dimensional impurities, the result was first obtained in~\cite{Takayanagi:2011zk}, while the $2+1$ dimensional example was studied in~\cite{Magan:2014dwa}. We compute the boundary entropies defined by entanglement, using the RT holographic prescription. We use the standard definition of the impurity entropy and compute the difference
\be
\label{impentropy}
S_{\rm imp} \ = \ S_{\rm E}[{\rm imp}] - S_{\rm E}[0]\,,
\ee
of the entanglement entropies in the presence and absence of an impurity. In particular, $S_{E}[{\rm imp}]$ refers to an impurity-containing subsystem of a bigger system.

In section~\ref{sec:1+1} we show that the AdS/BCFT configuration characterized by a constant tension brane in $AdS_3$ indeed corresponds to a conformal boundary condition: the impurity entropy computed from definition~(\ref{impentropy}) coincides with the thermodynamic entropy and is independent on whether we compute it in pure $AdS_3$ space, or in any other asymptotically $AdS_3$ static geometry.

The same constant tension brane in $AdS_4$, dual to a half-space boundary condition, considered in section~\ref{sec:2+1}, is no longer conformal. As follows from the thermodynamic entropy calculation, it corresponds to an irrelevant perturbation, since the defect entropy vanishes in the limit of zero temperature. The same is observed for the entanglement entropy, which renormalizes to zero for large sizes $R$ of the entanglement region. In section~\ref{sec:betabrane} we consider a finite temperature case and argue that the entanglement entropy renormalizes to one of the definitions of the entropy introduced in~\cite{Magan:2014dwa} for $T\sim 1/R$ (figure~\ref{fig:stripEntT}). At larger temperatures the effect of the impurity is expected to completely wash out due to thermal fluctuations. Interestingly, in the case of finite temperature, we observe a non-monotonic behavior of the entanglement entropy of the defect at $R\sim 1/T$.

Apart from constant tension brane solutions, in section~\ref{sec:betabrane} we also consider a brane with non-constant energy density. This type of solutions were first considered in~\cite{Erdmenger:2014xya}. These solutions realize a different type of boundary conditions in the BCFT. In particular, they may be thought as describing a single impurity on the circle, or a periodic array of impurities. With this finite density of impurities entanglement entropy shows a different behavior. It then looks like a relevant perturbation, as the entropy per impurity grows with size $R$ of the interval. Calculation of the entanglement entropy in this case involves a geometric transition. In comparison with the case with no impurities this transition occurs at a slightly different scale. At larger scales the impurities decouple and do not contribute to the to total entanglement entropy. The scale of the phase transition thus defines the screening radius of the impurities.

In concluding section~\ref{sec:conclusions} we discuss the results for the entropy obtained in section~\ref{sec:physics}. We interpret the behavior of the impurity entropy in terms of the RG flow. In the final part of section~\ref{sec:conclusions} we discuss open questions and future directions. We discuss some interesting observations made along the way. One of them is a holographic connection between solutions of section~\ref{sec:fininterval} and entanglement entropy of higher-dimensional strips, which might indicate an interesting duality of results in different dimensions. Second observation relates to the length of the branes considered in section~\ref{sec:betabrane} in $AdS_3$. Similarly to the geodesic lines, these branes also correctly compute the universal part of the entanglement entropy. We hope to elaborate on those observations in a future work.

\section{AdS/BCFT}
\label{sec:AdSBCFT}

\subsection{The model}

\begin{figure}[htb]
 \centering
 \includegraphics[width=0.4\linewidth]{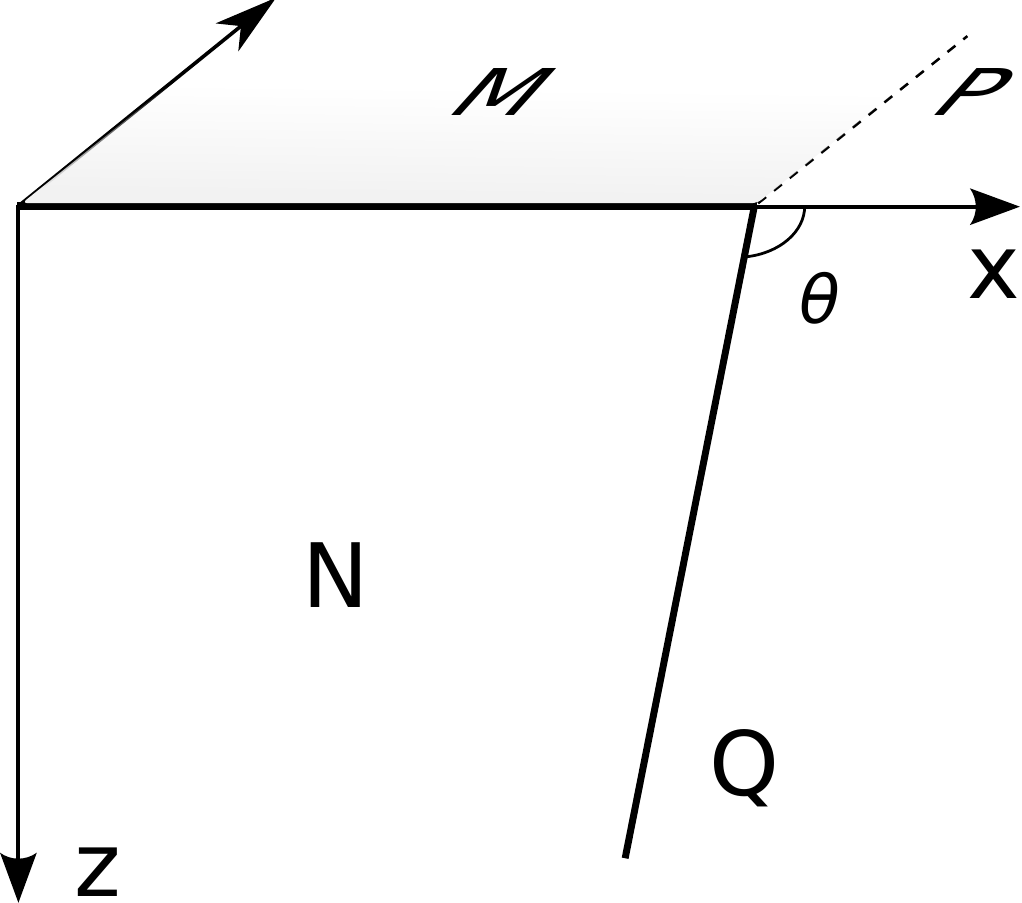}
 \caption{Illustration of boundary conditions in bulk space $N$ for a theory defined on boundary space $M$ with boundary $P$. Hypersurface $Q$ extends the boundary $P$ into the bulk $N$.}
 \label{fig:NMQP}
\end{figure}

We start by briefly reviewing the holographic dual description of a theory defined in a space with a boundary proposed by Takayanagi in~\cite{Takayanagi:2011zk}. Assume that we are interested in studying a theory in region $M$ of Minkowski space, which has boundary $P$. Presumably, the holographic dual of this theory is a gravity theory in space $N$, whose boundary includes $M$. Since $M$ itself has a boundary, $P=\partial M$, there should be another piece $Q$ of boundary $\partial N$ such that
\be
\partial N \ = \ M\cup Q\,, \qquad \text{and} \qquad \partial M\  =\  \partial Q \ = \ P\,.
\ee
This situation is illustrated in figure~\ref{fig:NMQP}.

Since the theory on $M$ with some boundary conditions on $P$ should completely define the quantum theory, the geometry of the dual gravity, including the geometry of $Q$ should be fixed by this data. As usual, boundary conditions on $M$ can be imposed via a deformation of the theory by a source term localized on $P$. It is in principle known, in top-down string theory constructions of holographic dual models, how to introduce such lower dimensional defects on the boundary. This is typically done considering intersecting D-branes~\cite{holodefects}. Most of the time the "defect" branes are considered in the probe approximation, which means that they are introduced in a fixed gravity background and the backreaction is neglected. The geometry of the embedding is controlled by an effective action, which depends, besides other things, on their world-volume metric induced from the background. The induced metric determines the embedding profile of a probe brane in the bulk. It is found by a minimization of the action.

The construction of Takayanagi, reviewed here, ignores the glorious details of the top-down models. This approach is specifically justified in the study of lower-dimensional examples of holography, such as the AdS$_3$/CFT$_2$ correspondence, where universal geometric features map onto well-known properties of 2D CFTs. The approach is nevertheless based on the idea that the profile of the boundary $Q$ should be determined by dynamical equations -- the variational principle. Consequently, boundary $Q$ is introduced in the bulk gravity action through a $Q$-localized term.

The complete action of the gravity theory dual to a $d$-dimensional theory on $M$ is a sum of various pieces:
\begin{multline}
\label{action}
I =  \frac{1}{2\kappa}\int_N \d^{d+1}x\, \sqrt{-g}(R-2\Lambda)+\frac{1}{\kappa}\int_Q \d^{d}x\, \sqrt{-h}(K-\Sigma) +\frac{1}{\kappa}\int_M\d^dx\, \sqrt{-\gamma}(K^{(\gamma)}-\Sigma^{(\gamma)})+ \\
+ \int_Q\d^dx\, \mathcal{L_{\rm mat}}  + \Delta I\,.
\end{multline}
Here $\kappa=8\pi G$ is the gravitational coupling constant, $g_{\mu\nu}$ is the bulk metric, $h_{ab}$ and $\gamma_{ij}$ are induced metrics on $Q$ and $M$, $K$ and $K^{(\gamma)}$ are corresponding traces of the extrinsic curvature, $\Sigma$ and $\Sigma^{(\gamma)}$ are tensions of $Q$ and $M$ respectively. $\mathcal{L}_{\rm mat}$ is a Lagrangian of possible matter fields on $Q$. $\Delta I$ is the part of the action that contains possible counter-terms and contact terms, localized on $P$. They do not affect the bulk dynamics, but are introduced to make the action finite. We will not consider their exact form in this paper (see~\cite{AdS/BCFT2,Magan:2014dwa} for more details). In principle, bulk theory can contain additional fields with corresponding boundary terms acting as sources for boundary operators, \emph{e.g.}~\cite{QHE}. For simplicity we restrict to the minimal sector, only containing gravity, as in the original proposal.

As usual, the variation of the action yields equations of motion up to a boundary term. One should choose some boundary conditions to completely define the model. As far as the $M$-part is concerned, the most common choice is the Dirichlet boundary conditions, which does not allow metric on $M$ to vary. For $Q$ one would like the induced metric to be determined dynamically, so it is proposed to use the Neumann boundary conditions, that is to fix the stress-energy tensor, which is the canonically conjugate quantity with respect to the metric variation. In terms of the above action, this amounts to imposing a condition similar to the Israel junction condition,
\be
\label{Neumann-metric}
K_{ab} - K h_{ab} = \kappa T_{ab}-\Sigma h_{ab}\,.
\ee
This is a $Q$-projected equation, in terms of the induced metric $h_{ab}$, with $K_{ab}$ being the pullback of the extrinsic curvature on $Q$ ($K$ being its scalar). The right hand side of the equation is the stress-energy tensor of the matter Lagrangian, in units of $\kappa$. In accordance with equation~(\ref{action}), the constant energy density piece (surface tension $\Sigma$, or equivalently, cosmological constant on $Q$) is made explicit. $T_{ab}$ may also contain the contribution from the counter terms.

Given $T_{ab}$ (and $\Sigma$) equation~(\ref{Neumann-metric}) can be solved to find the induced metric $h_{ab}$, and hence the profile of the boundary $Q$ in the bulk. In the remainder of this section we will review some solutions to equations~(\ref{Neumann-metric}) for  $d=1+1$ and $2+1$, relevant for the later discussion. Some other solutions are presented in appendix~\ref{appendix}. Although we work in the Poincar\'e coordinates, some examples are also generalized to the case of global coordinates in the appendix~\ref{global}.

\subsection{Relevant examples}

In this section we discuss some basic examples of the AdS/BCFT construction of~\cite{Takayanagi:2011zk}. We mostly discuss known solutions, though some new generalizations are presented in appendix~\ref{appendix}. More recent examples can be found in~\cite{Chu:2017aab,AdSBCFTexamples}. Moreover, authors of~\cite{TopDownExamples,Nagasaki:2011ue} discuss similar top-down and bottom-up constructions of the holographic duals of BCFTs. In~\cite{Chu:2017aab,AdS/BCFT3} boundary conditions alternative to~(\ref{Neumann-metric}) are considered. We will mention some of the the results of those papers in section~\ref{sec:2+1}.

The basic gravity backgrounds we will consider are those of the empty anti de Sitter space given by the metric
\be
\label{AdS3}
ds^2 \ = \ \frac{L^2}{z^2}\left(-dt^2 + dx^2 + \sum\limits_{i=0}^{d-2}dy_i^2 + dz^2\right),
\ee
and the asymptotically $AdS_{d+1}$ Schwarzschild black holes
\be
\label{blackhole}
ds^2 \ = \ \frac{L^2}{z^2}\left(- f(z)dt^2 + dx^2 + \sum\limits_{i=0}^{d-2}dy_i^2 + \frac{dz^2}{f(z)}\right).
\ee
The possible transverse directions labeled by coordinates $y_i$ are less relevant in this work. The blackening factor $f(z)$ of the black hole is
\be
f(z) \ = \ 1- \frac{z^{d}}{z_h^d}\,,
\ee
where $z_h$ is the horizon ``radius'' of the black hole.

Thermal $AdS$ is another geometry that we will consider. It corresponds to
\be
\label{thermal}
ds^2 \ = \ \frac{L^2}{z^2}\left(- dt^2 + h(z)dx^2 + \sum\limits_{i=0}^{d-2}dy_i^2 + \frac{dz^2}{h(z)}\right),
\ee
where $h(z)=1-z^d/z_0^d$. In three dimensions all the above geometries are related to each other, and to any other asymptotically $AdS_3$ solution, by (large) diffeomorphisms.

\subsubsection{Prime example. Holographic dual of half-space}
\label{sec:basic}

Let us first study the case of half-space boundary condition, which without loss of generality we set as
\be
\label{halfspace}
P:\quad x\ = \ 0\,,
\ee
in terms of coordinate parametrization~(\ref{AdS3}). By symmetry of the boundary condition we can parameterize the profile of the boundary $Q$ as
\be
\label{line}
Q: \quad x\ = \ x(z)\,.
\ee

The simplest boundary conditions one can choose in~(\ref{Neumann-metric}) is the one in which only the surface tension term $\Sigma$ is present, that is $T_{ab}=0$. In pure $AdS_3$~(\ref{AdS3}) equations~(\ref{Neumann-metric}) has a simple solution~\cite{Takayanagi:2011zk},
\be
\label{Qprofile0}
x(z) = z\cot\theta\,, \qquad \text{where} \qquad \cos\theta=L\Sigma\,.
\ee
In other words, $Q$ is a straight line in coordinate parametrization~(\ref{line}). Surface tension $\Sigma$ controls the angle at which line~(\ref{Qprofile0}) intersects the boundary $M$ (see figure~\ref{fig:NMQP}). We define $\theta$ as the angle external to region $N$ encoding physics in $M$.

The induced metric on $Q$ is a slice of $AdS_2$:
\be
ds^2_{\rm ind} \ = \ \frac{L^2}{z^2}\left(-dt^2 + \frac{dz^2}{\sin\theta^2}\right).
\ee
One can see that the case $0\leq\theta< \pi/2$ corresponds to positive tension $\Sigma>0$. Tension is negative for $\pi/2<\theta\leq \pi$. In both cases tension is bounded $|\Sigma|\leq 1/L$.

Provided boundary condition~(\ref{halfspace}) and metric~(\ref{AdS3}) it is straightforward to generalize this solution to arbitrary boundary dimension $d$. (See~\cite{AdS/BCFT2} for more details.) Angle $\theta$ in this case is related to the tension through
\be
\label{TTrel}
(d-1)\cos\theta \ = \ \Sigma L\,.
\ee
A $d=5$ example of a simple $D3$-$D5$ brane system realizing such a profile was proposed in~\cite{Nagasaki:2011ue}.

\subsubsection{Half-space in \texorpdfstring{$AdS_3$}{AdS3} at finite temperature}

\begin{figure}
 \centering
 \includegraphics[width=0.5\linewidth]{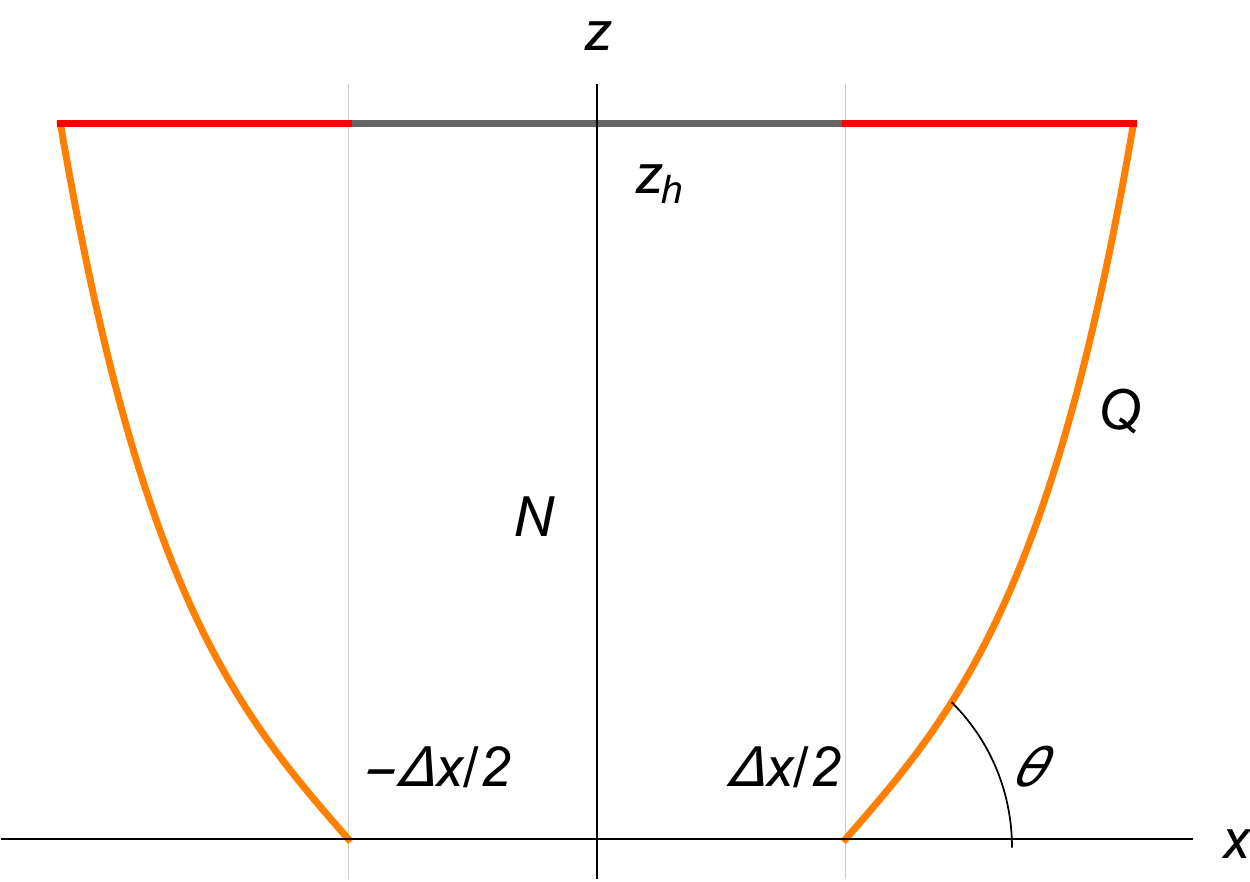}

 \caption{Profile of the boundary $Q$ in the case of the BTZ black hole. Red regions show the "shadows" of the boundaries $Q$ on the horizon, which contribute to the boundary entropy.}
 \label{fig:BTZ}
\end{figure}

Boundary condition~(\ref{line}) with $T_{ab}=0$ can be also solved at finite temperature in $AdS_3$~\cite{Takayanagi:2011zk}. In the $d=1+1$ version of metric~(\ref{blackhole}) one finds
\be
\label{QBTZ}
x(z)=z_h\,{\rm arcsinh}\left(\frac{z}{z_h}\cot\theta\right)\,.
\ee
The profile of $Q$ is demonstrated on figure~\ref{fig:BTZ}, where two impurities and, consequently, two branches of $Q$ are shown. Angle $\theta$, again, is the angle at which $Q$ crosses the boundary at $z=0$, external to subspace $N$. For $z\to 0$, one reproduces the result of empty $AdS$~(\ref{Qprofile0}). For $z\to z_h$ the profile enters the horizon at a finite angle, $x=x_0+\Delta x+\cos\theta'(z-z_h)$, $\cot\theta'=\cos\theta$.

It turns out to be difficult to generalize these solutions to higher dimensions keeping the simplest scenario of only surface tension. The problem is that~(\ref{Neumann-metric}) is a tensor equation and it becomes difficult to make it self-consistent with only few parameter functions. (See~\cite{Nozaki:2012qd} for further discussion.) In~\cite{AdS/BCFT3} a simpler scalar equation was proposed to make the problem solvable also in higher dimensions. Here, instead, we will consider non-trivial matter content on $Q$, such that $T_{ab}\neq 0$.

\subsubsection{Finite size interval in \texorpdfstring{$AdS_3$}{AdS3}}
\label{sec:TAdS}

Solutions obtained in the previous two examples correspond to semi-infinite intervals $M$. In the boundary CFT one can apply a compactifying conformal transformation, mapping the system onto a finite interval $M$. This transformation changes $AdS_3$ metric~(\ref{AdS3}) to that of the thermal AdS, equation~(\ref{thermal}) for $d=2$. Under this transformation equation~(\ref{Neumann-metric}) transforms covariantly, so there is also a solution with $T_{ab}=0$,
\be
\label{QTAdS}
x(z) \  = \ z_0 \arctan\left(\frac{z\cos\theta}{z_0\sqrt{h(z)-\cos^2\theta}}\right).
\ee

In order for metric~(\ref{thermal}) to be non-singular one has to assume that $x$ coordinate is compact with periodicity $2\pi z_0$ and $z$ is bounded: $z\leq z_0$. The space of $x$ and $z$ coordinates has the shape of a cigar whose tip is at $z=z_0$. Boundary $Q$ described by equation~(\ref{QTAdS}) has a U-shaped profile anchored at the opposite points on the $x$-circle ($x=0$ and $x=\pi$ in case of the above solution). It has a turning point at $z=z_\ast$, such that $h(z_\ast)=\cos^2\theta$.

\subsubsection{Intervals of arbitrary size}
\label{sec:fininterval}

In the thermal $AdS_3$ example above the boundary interval $M$ is fixed to be a half of the space. Some extra parameters are necessary in order to be able to vary the relative size of the interval. One construction was proposed in~\cite{Erdmenger:2014xya}. Let us briefly review and generalize it.

Let us consider a more general set of boundary conditions, with $T_{ab}$ in equation~(\ref{Neumann-metric}) different from zero. In other words, we will consider a local conformal transformation, introducing a non-constant energy-density. We can ask what kind of stress-energy tensor on $Q$ is compatible with asymptotically $AdS$ metric~(\ref{AdS3}) and~(\ref{blackhole}). For the half-space boundary condition~(\ref{halfspace}) and parametrization~(\ref{line}) the form of the stress-energy is
\be
\label{Tab}
T_{ab} \ = \ \left(\begin{array}{ccc}
                     - \varepsilon(z) h_{tt} && \\
                      & p_{y}(z)\delta_{ij}h_{yy} &  \\
                      && p_z(z)h_{zz}
                   \end{array}\right),
\ee
where, in the case of a general boundary space-time dimension $d$, $\delta_{ij}$ is a $(d-2)\times(d-2)$ unit matrix referring to transverse dimensions and
\begin{eqnarray}
\varepsilon \left( z\right)  &=&\frac{\left( 1-d\right) x^{\prime }+\left(
1-d\right) x^{\prime 3}+zx^{\prime \prime }}{L\left( 1+x^{\prime 2}\right)
^{3/2}}, \\
p_{y}\left( z\right)  &=&\frac{\left( d-1\right) x^{\prime }+\left(
d-1\right) x^{\prime 3}-zx^{\prime \prime }}{L\left( 1+x^{\prime 2}\right)
^{3/2}} \\
p_{z}(z) &=&\frac{\left( d-1\right) x^{\prime }}{L\sqrt{1+x^{\prime 2}}}.
\end{eqnarray}
If one imposes some additional physical conditions on $T_{ab}$, one can fix the shape of the function $x(z)$.

\begin{figure}[htb]
\centering
 \includegraphics[width=0.5\linewidth]{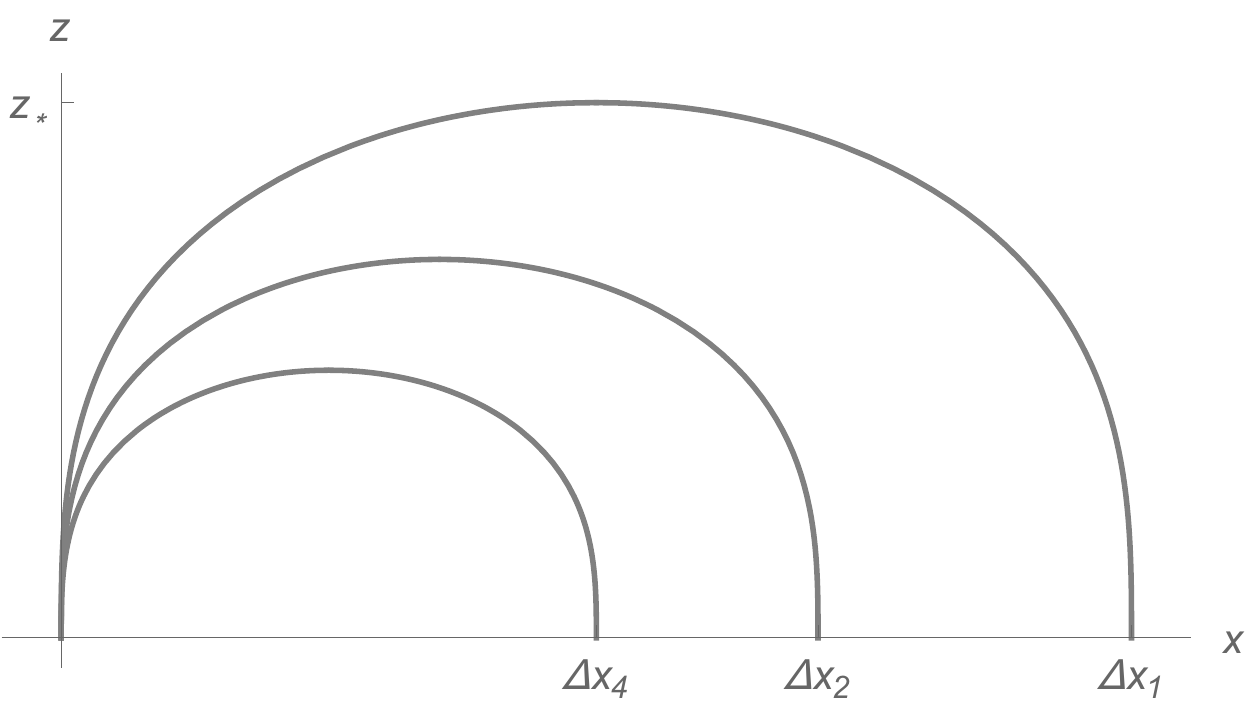}
 \caption{Set of reconnecting profiles studied in~\cite{Erdmenger:2014xya} for conformal energy-momentum tensor on $Q$.}
 \label{fig:betaplots}
\end{figure}

In the $d=1+1$-dimensional example of reference~\cite{Erdmenger:2014xya} the physical condition imposed was the equation of state $\varepsilon/p={\rm const}$, where $p\equiv p_z$. The specific case $\varepsilon=p$ is the case of conformal $T^{ab}$. This condition directly generalizes to an arbitrary dimension:
\be
\label{conformalT}
\tr T_{ab} \ = \ - \varepsilon + p_z + \sum\limits_{i}^{d-2}p_{y_i} \ = \ 0\,.
\ee
The above equation becomes a second order equation on $x'(z)$, from which the profile of $Q$ can be determined. In dimension $d$ this equation can be reduced to
\be
\label{betaequation}
zx'' \ = \ dx'(1+x'^2)\,.
\ee
In $d=2$ there is a nice representation of the solution to this equation in terms of an incomplete Euler beta function:
\be
\label{beta}
x(z) \ = \  \frac{1}{4}B_{z^4}\left(\frac{3}{4},\frac{1}{2}\right).
\ee
The solution has a scaling symmetry $z\to \lambda z$, $x\to \lambda x$: if $x(z)$ is a solution, then $\lambda x(\lambda^{-1}z)$ is also a solution. Parameter $\lambda$ sets the length of the interval $M$. It is in fact the value of $z$ at the turning point. In figure~\ref{fig:betaplots} we demonstrate the profiles for several values of $\lambda$. The length of the intervals is given by $\lambda B(3/4,1/2)$.

One can make an interesting observation regarding equations~(\ref{betaequation}). If one considers the RT problem of computing the area of a minimal surface in $AdS_{d+2}$ anchored at a $d+1$-dimensional infinite strip of finite width, then equation~(\ref{betaequation}) would provide a solution of that problem.

\subsubsection{Finite temperature example in higher dimension}
\label{sec:MMS}

In the previous example, we have provided a generalization of the AdS/BCFT problem on an infinite strip with conformal $T_{ab}$ to arbitrary dimensions. This was done observing that conformal constraint uniquely fixes the form of the surface $Q$. This generalization works either for zero and for non-zero temperature $T$. One can also impose different physical conditions and extract other profiles of $Q$. In reference~\cite{Magan:2014dwa} a fluid condition was imposed. This conditions requires that the stress-energy tensor is that of a perfect fluid, which means requiring
\be
\label{fluidBC}
p_z \ = \ p_{y_i}\,,
\ee
for all $i$. Again, such a condition can be solved in any dimension for the half space boundary condition~(\ref{halfspace}). The solution has the form
\be
\label{MMSprofile}
x(z) \ = \  \cot\theta\int^{z} \frac{d q}{\sqrt{f(q)}} \,,
\ee
where $f(z)$ is the blackening factor of the metric~(\ref{blackhole}). We remind that in dimension $d$ angle $\theta$ is related to the tension via~(\ref{TTrel}). In $d=2$, the result can be expressed in terms of a hypergeometric function.

\begin{figure}
 \centering
 \includegraphics[width=0.5\linewidth]{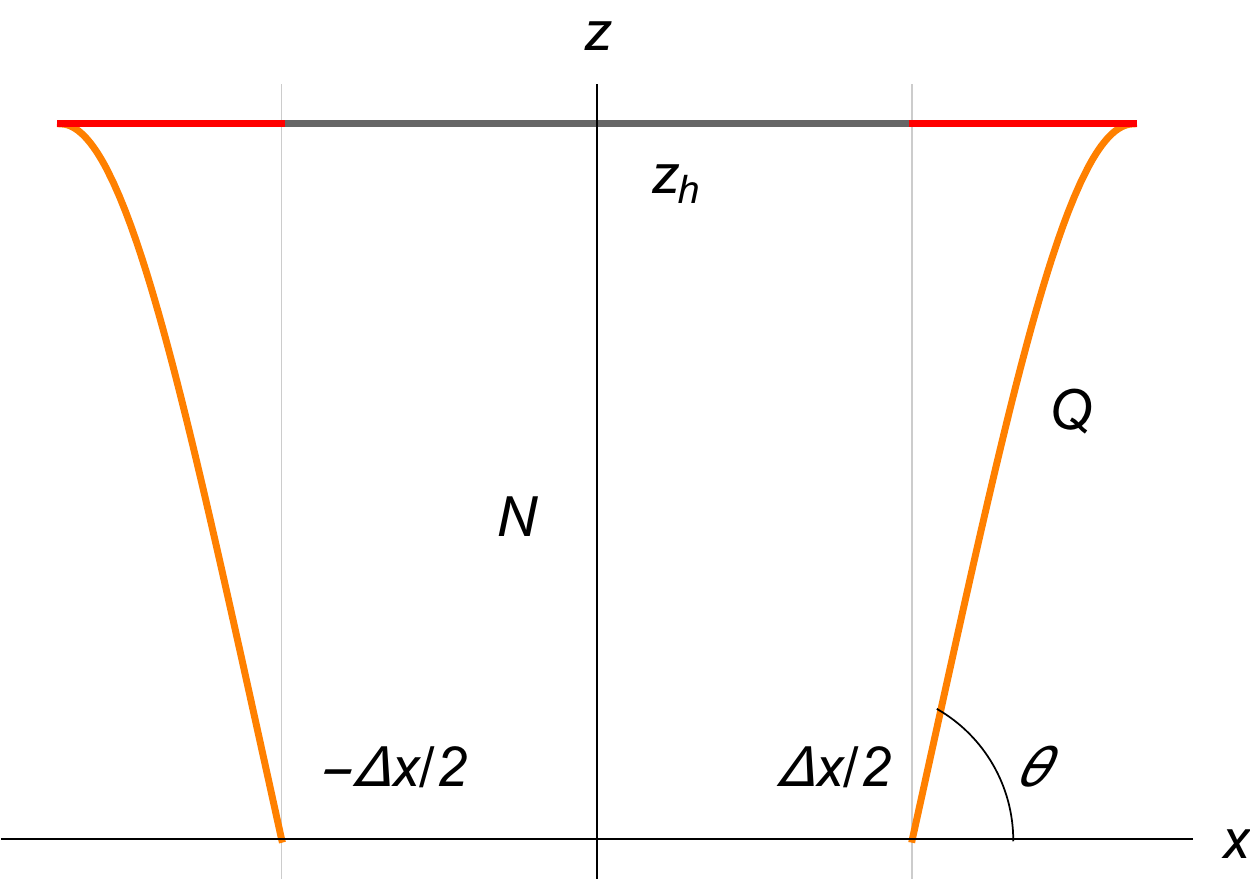}

 \caption{Profile of the boundary $Q$ in the case of $AdS_4$ black hole with boundary condition~(\ref{fluidBC}). Red regions show the part of the horizon (shadows of $Q$), which contributes to the boundary entropy.}
 \label{fig:MMS}
\end{figure}

A setup with two boundaries defined by solution~(\ref{MMSprofile}) in $AdS_4$ space is shown on figure~\ref{fig:MMS}. In contrast to solutions with conformal $T_{ab}$ the fluid-like profiles do not reconnect to the boundary. In $T\to 0$ limit they reduce to solutions~(\ref{Qprofile0}).

\section{Basic physics of quantum impurities}
\label{sec:physics}

In this section we start discussing the properties of the solutions to the AdS/BCFT problem, reviewed in the previous section, from the point of view of impurity/defect physics. One interesting physical quantity that characterizes the nature of a defect is its entropy, a measure of degrees of freedom associated to it. In $1+1$-dimensional case the entropy has some special properties which we will now demonstrate using the geometric picture.

\subsection{Thermodynamic entropy}
\label{sec:TEntropy}

Let us consider a finite-temperature $1+1$-dimensional theory on an interval of size $\Delta x$. We assume that the interval is bounded by two impurities, whose effect is introduced by special boundary conditions~(\ref{Neumann-metric}). The system is assumed to be described by action~(\ref{action}). Diagrammatically, the setup corresponds to the one shown on figure~\ref{fig:BTZ}.

Thermodynamics of this system can be computed from the free energy, which is given by the Euclidean action computed on the solution illustrated by figure~\ref{fig:BTZ}. The original calculation was performed in~\cite{Takayanagi:2011zk}. Some additional details can be found in the appendix of~\cite{Magan:2014dwa}. After a subtraction of appropriate counterterms the regularized Euclidean action can be cast in the form
\be
-\beta F \ = \ I_{E} \ = \  I_{\rm bulk} + 2 I_{\rm bry}\,,
\ee
with explicit contributions given by
\begin{eqnarray}
I_{\rm bulk} &  =  & - \frac{L}{8G}\,\frac{\Delta x}{z_h}\,, \\
I_{\rm bry} & = &   - \frac{L}{4G}\,{\rm arcsinh}({\cot\theta})\,.
\label{Ibdry1}
\end{eqnarray}
This separation of the total action into the bulk and boundary pieces is indeed sensible, since the bulk term is extensive, while the boundary term depends only on the impurity parameter $\Sigma$, through the geometric angle $\theta$.

Consequently, the entropy can be computed by taking an appropriate derivative of the free energy
\be
\label{BTZentropy}
S = -\frac{\partial F}{\partial T}= \frac{L}{4G}\frac{\Delta x}{z_h}+ \frac{L}{2G}\,{\rm arcsinh}({\cot\theta})\,.
\ee
One can notice that both contributions are consistent with the Bekenstein-Hawking scaling: while the bulk term gives the standard entropy proportional to the size of the boundary system through the Bekenstein-Hawking entropy density, the boundary contribution does not have a ``size", but its entropy has a geometric interpretation as the Bekenstein-Hawking coefficient times the area of the black hole horizon immediately below the $Q$-brane (see figure~\ref{fig:BTZ}). Here we refer to this part of the horizon as to the shadow of $Q$. Hence we obtain
\be
\label{SQ}
S_Q \ = \ \frac{c}{6}\,{\rm arcsinh}({\cot\theta})\,,
\ee
where the Brown-Henneaux formula was used to express the gravity parameters $L$ and $G$ in terms of the central charge.

Another finite-temperature example reviewed in the previous section was the solution with fluid-like stress-energy tensor $T_{ab}$, (\ref{Tab}) subject to conditions~(\ref{fluidBC}), in equation~(\ref{Neumann-metric}). An analogous thermodynamical computation  in the configuration shown on figure~\ref{fig:MMS} (infinite strip of width $\Delta x$ in $d=2+1$) was performed in~\cite{Magan:2014dwa}. The result of reference~\cite{Magan:2014dwa} can be cast in the form
\be
\label{Snew}
S\equiv S_{\rm bulk} + 2S_{\rm bry} = -\frac{\partial F}{\partial T_H}= \frac{16\pi^2}{9}\,cT^2\Delta x\Delta y+ 2\,\frac{16\pi}{9}\,cT\Delta y\cot\theta\,.
\ee
Here $\Delta y$, is the (infinite) length of the strip, so that the finite quantity is the line entropy density. For the central charge in $3+1$-dimensional gravity we use
\be
c \ = \ \frac{L^2}{4G}\,.
\ee
We remind also that in $AdS_4$ case angle $\theta$ is connected with the brane tension through $\Sigma L \ = \ 2\cos\theta$. Hence, the boundary entropy, which follows from the thermodynamical calculation in~\cite{Magan:2014dwa} is
\be
\label{Bent2d}
\frac{S_T}{\Delta y} \ = \ \frac{16\pi}{9}\,cT\cot\theta\,.
\ee

As it was observed in~\cite{Magan:2014dwa} entropy~(\ref{Bent2d}) does not respect the Bekenstein-Hawking (BH) scaling, that is, it is not equal to the area of the shadow of the bounding surface $Q$ on the horizon (marked red on figure~\ref{fig:MMS}) times the correct BH factor. Instead, analog of the BH entropy for the boundary would be
\be
\label{BHentropy}
\frac{S_{\rm BH}}{\Delta y} \ = \ \frac{\sqrt{\pi}L^2}{4Gz_h}\,\frac{\Gamma(4/3)}{\Gamma(5/6)}\cot\theta \ = \ \frac{4\pi^{3/2}}{3}\,\frac{\Gamma(4/3)}{\Gamma(5/6)}\, c T\cot\theta \,.
\ee
This entropy was also derived in~\cite{Magan:2014dwa} from the consistency of thermodynamic relations of the fluid, described by $T_{ab}$ on $Q$. One can check, in fact that the two results differ by approximately 5\%. Below, in section~\ref{sec:betabrane}, we will show that equation~(\ref{BHentropy}) appears to be consistent with the entanglement entropy of the defect in the thermal state.

\subsection{Impurity entropy from entanglement}

Now let us discuss the entropy associated with the boundary using another probe, the entanglement entropy.

In holography entanglement entropy can be calculated using the Ryu-Takayanagi formula~\cite{RT}. If a system, which has a gravity dual, is partitioned in a disjoint union $A\cup B$ then the entanglement entropy of a part, say $A$, can be computed as the area of a minimal surface $\gamma$ in the gravity bulk, such that $\partial \gamma = A$:
\be
\label{RTformula}
S_{\rm E}[A] \ = \ \min\limits_{\gamma|A=\partial \gamma}\frac{{\rm Area}[\gamma]}{4G}\,.
\ee
In the case of three-dimensional gravity, the minimal surfaces are the geodesic lines connecting points at the boundary and the entropy is proportional to their length.

\begin{figure}[h]
 \begin{minipage}{0.45\linewidth}
  \includegraphics[width=\linewidth]{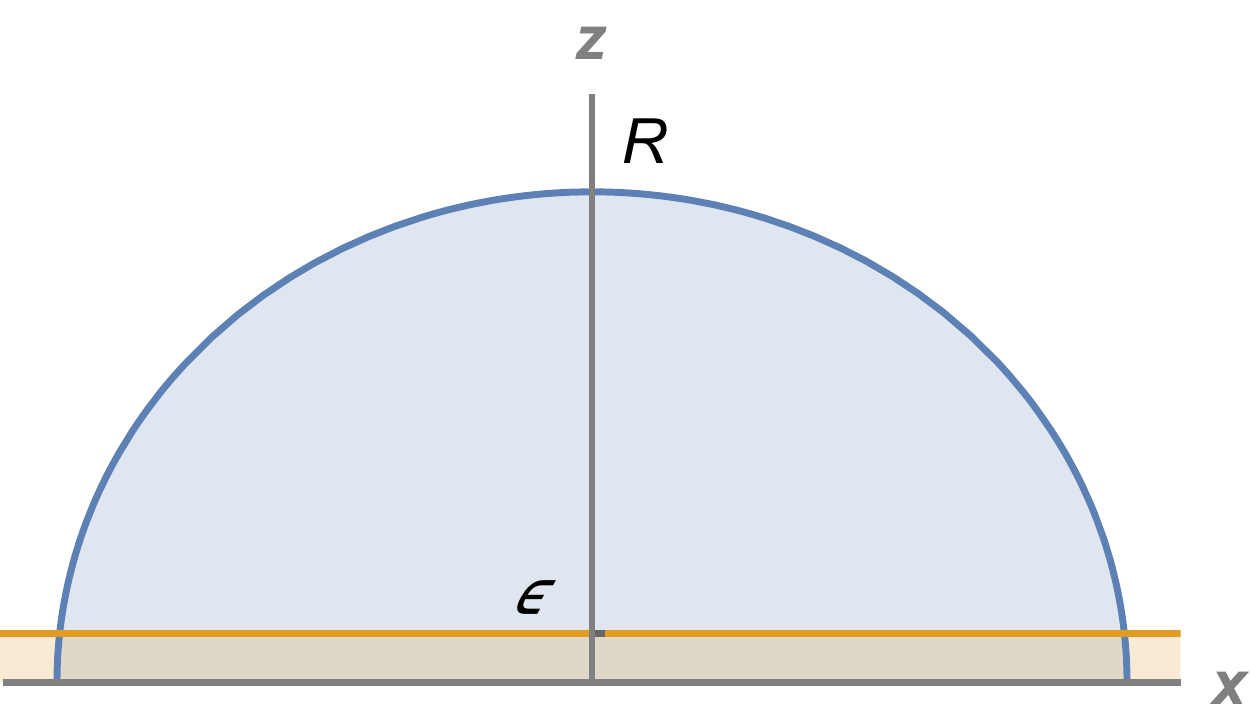}
 \end{minipage}
\hfill{
\begin{minipage}{0.45\linewidth}
  \includegraphics[width=\linewidth]{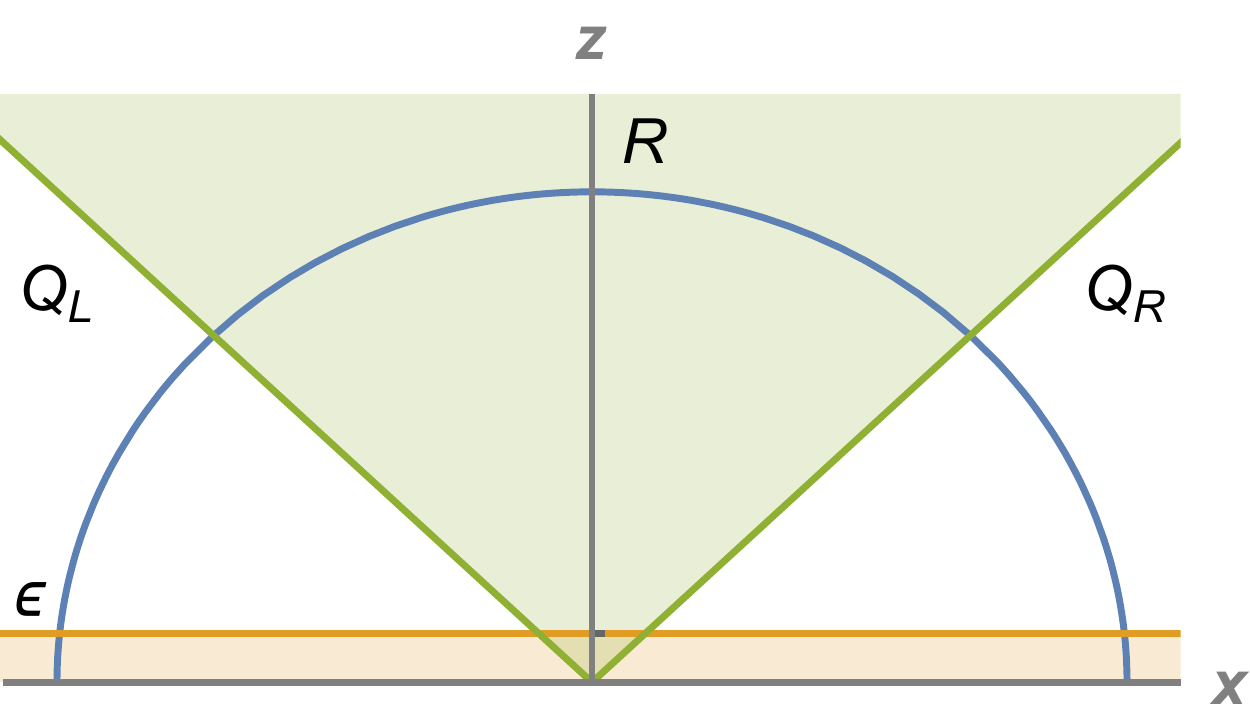}
 \end{minipage}
}
 \caption{Holographic definition of the impurity entropy. The shaded region in the bottom is excised due to the cutoff $z=\epsilon$. The corresponding piece of the curve has an infinite area there. Left: RT (minimal area) curve in the absence of impurity. Right: Impurity in the bulk is represented by two surfaces $Q_L$ and $Q_R$. The shaded space between them is excised and two surfaces must be identified, so that the impurity creates angle deficit $2(\pi-\theta)$. The illustrated case corresponds to $\theta>\pi/2$, or negative tension $\Sigma\leq 0$. In the case of positive tension, $\theta<\pi/2$, there is a proficit of angle.}
 \label{fig:impent}
\end{figure}

Let us imagine an infinite line containing a finite interval with an impurity at the center. In the absence of the impurity, the entanglement entropy of the interval with respect to its infinite complement is proportional to the length of the geodesic line (as long as the $d=1+1$ example is discussed) in the AdS bulk anchored at the endpoints of the interval, as shown on figure~\ref{fig:impent} (left).

When impurity is introduced, it will create a defect, or a special boundary condition on the surface $Q$ in the bulk. In fact, surface $Q$ should be two-sided and we have to impose boundary conditions ``across'' it, exactly as in the Israel junction condition~\cite{Israel}. This is well explained in reference~\cite{Erdmenger:2014xya}. Effectively one has to ``glue'' together two pictures like the ones shown on figure~\ref{fig:NMQP}, so that the part of the space between them is excised and the boundaries $Q$ are identified, as on figure~\ref{fig:impent} (right).

Put differently, the defects will create a deficit, or a proficit of angle, since a part of the space is excised/added. So if we compute the entanglement entropy of an interval with the impurity in the middle, one should through away or add, part of the length of the geodesic corresponding to the excised or added part of the space. Consequently, one defines the impurity entropy as in equation~(\ref{impentropy}), subtracting the entropy in the presence and in the absence of the impurity.

Note that boundary condition~(\ref{Neumann-metric}) needs to be rectified to take into account the fact that $Q$ is two-sided. In particular, we can write $T_{ab}$ on $Q$ as
\be
T_{ab} \ = \ T_{ab}^R - T_{ab}^L\,,
\ee
where left and right contributions are defined in terms of the left and right extrinsic curvatures and induced metrics. Since we are considering symmetric configurations, the left and right contributions are equal in the magnitude, but opposite in the sign, so the only effect of this rectification would be a renormalization of $T_{ab}$ by a factor of two. This will also affect by a factor of two relation~(\ref{TTrel}) between $\Sigma$ and $\theta$.

In generic $1+1$-dimensional theories the entanglement entropy associated with an impurity are functions of a characteristic energy scales (like temperature, or the length of the entanglement interval). At low energies (small temperature, large interval) the entropy runs to a fixed point value, controlled by a $1+1$-dimensional CFT. In a $1+1$-dimensional CFT the value of the impurity entropy should be equal to the thermodynamic entropy discussed above~\cite{CC}. In the following examples we will demonstrate this property.

\subsubsection{Constant energy density in 1+1 dimensions}
\label{sec:1+1}

\begin{figure}
 \begin{minipage}{0.45\linewidth}
   \includegraphics[width=\linewidth]{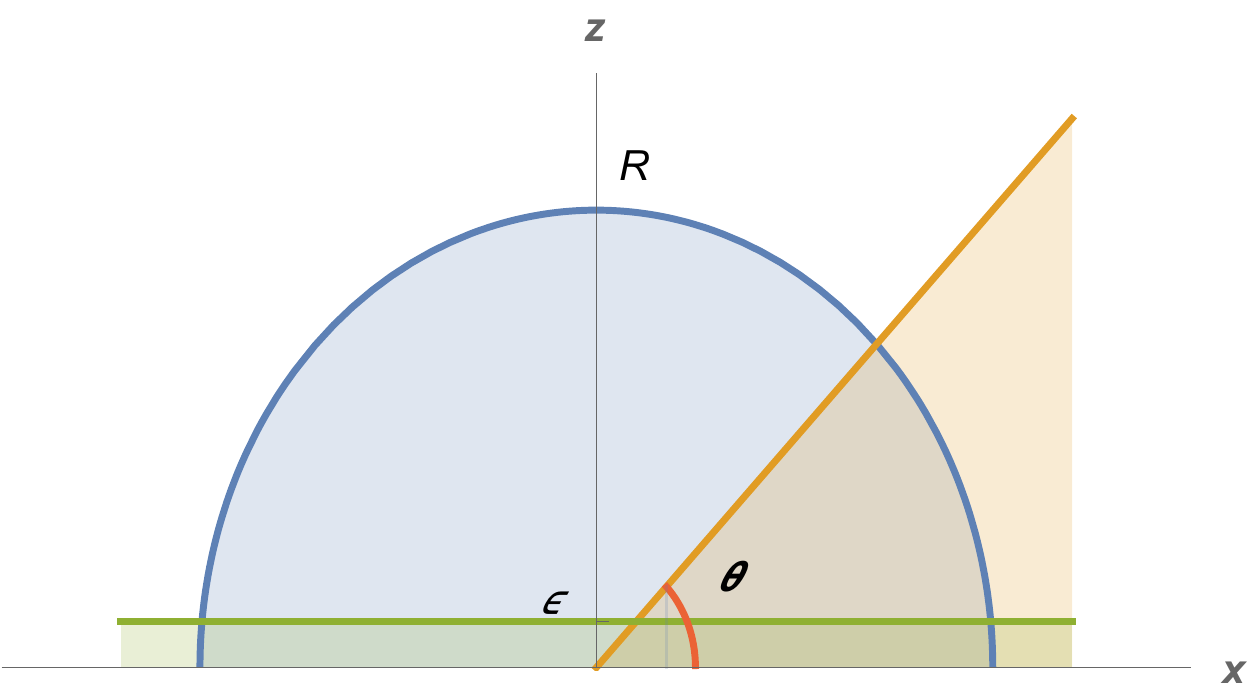}
 \end{minipage}
\hfill{
\begin{minipage}{0.45\linewidth}
   \includegraphics[width=\linewidth]{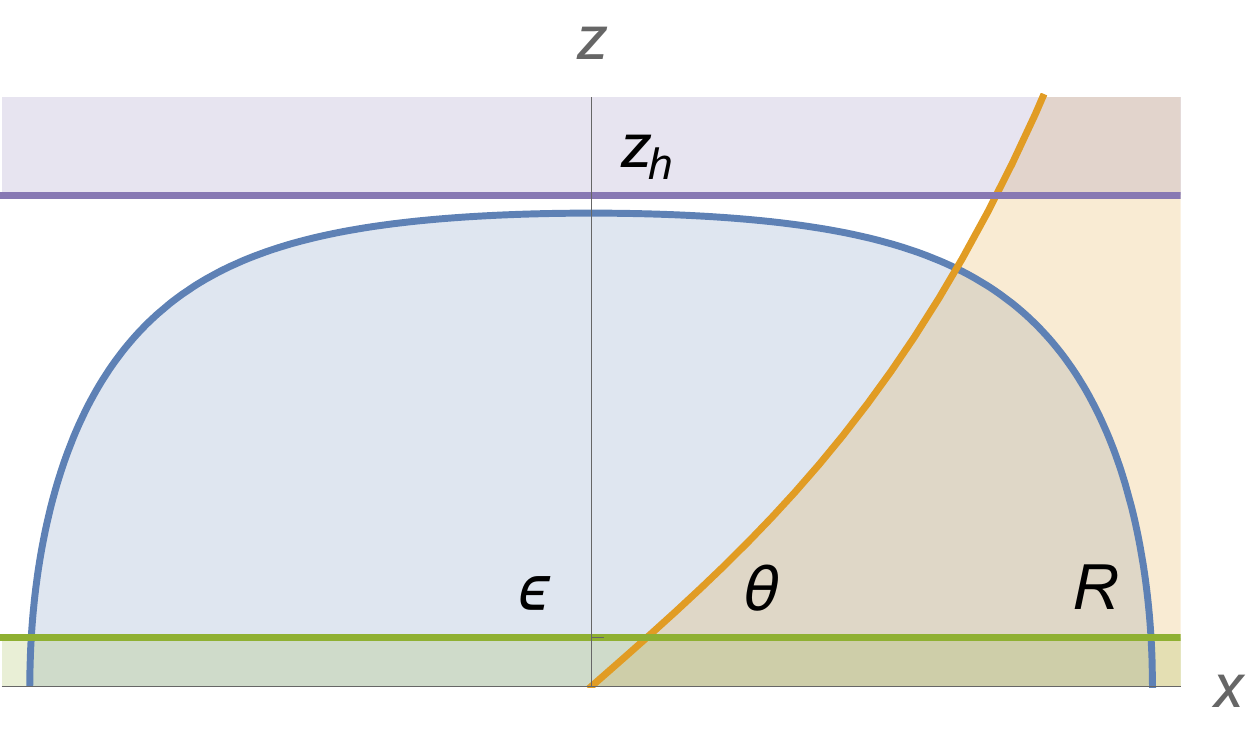}
 \end{minipage}
}
 \caption{Calculation of entanglement entropy in the case of a single impurity introduced by a constant tension brane in $AdS_3$ in the cases of zero (left) and finite (right) temperature.}
 \label{fig:3dT}
\end{figure}

As have already been mentioned, in the case of a $d=1+1$ system, the minimal (RT) surfaces are geodesic lines and the area is their length. In the absence of impurities, the entanglement entropy of an interval of length $2R$ is given by the length of the geodesic connecting two endpoints of the interval through the bulk. In Poincar\'e coordinates~(\ref{AdS3}) the geodesics are simply the semicircles of radius $R$ connecting the endpoints. One finds that
\be
\label{SEinterval}
S_{\rm E}(R) \ = \ \frac{L}{4G_N}\log\left(\frac{R+\sqrt{R^2-\epsilon^2}}{R-\sqrt{R^2-\epsilon^2}}\right).
\ee
Since $Q$ is two-sided one can divide this result into left and right halves and compute the correction due to the angle deficit created by the impurity for each half separately in accordance with figure~\ref{fig:3dT} (left). This is equivalent to computing the entropy of an interval on a semi-infinite line with an impurity at the origin.

Thus, for a semi-infinite line one finds
\be
 \frac{L}{2}\log\left(\frac{R+\sqrt{R^2-\epsilon^2}}{R-\sqrt{R^2-\epsilon^2}}\cot^2\frac{\theta}{2}\right),
\ee
where one needs to keep only singular and finite terms in the limit of the cutoff $\epsilon\to 0$. Hence, the entropy can be cast in the form
\be
\label{BCFTEE}
S_{\rm E} \ = \ \frac{c}{6}\log\frac{2R}{\epsilon} + \log g\,,
\ee
where $g$ is a $R$-independent function characterizing the impurity:
\be
\label{D2Entropy}
\log g \ = \ \frac{c}{6}\log\cot\frac{\theta}{2} \ = \ S_Q\,.
\ee
As a consistency check, this entropy vanishes, when $\theta=\pi/2$.

Written in this form, equation~(\ref{BCFTEE}) is a well-known universal result in BCFT. Function $g$ measures the boundary degrees of freedom and in particular, $\log g = S_{\rm imp}$. Impurity entropy matches contribution~(\ref{SQ}) calculated in the previous section. This result was outlined by Takayanagi in his AdS/BCFT paper~\cite{Takayanagi:2011zk}. In the holographic context it was perhaps first discussed in~\cite{Azeyanagi:2007qj}, see also~\cite{TopDownExamples}.

Impurity entropy~(\ref{D2Entropy}) is a number independent from the size of the interval. It is characterized entirely by the boundary condition (angle $\theta$, or tension $\Sigma$) and a CFT in question (central charge). This makes it a well-defined boundary quantity, a fixed-point of the RG flow. Indeed, one can do the same calculation, now in the finite temperature, asymptotically $AdS_3$ black hole background (BTZ black hole~\cite{BTZ}) given by metric~(\ref{blackhole}) for $d=2$. The relevant configuration is shown on figure~\ref{fig:3dT} (right).

The profile of the surface $Q$ in this case is given by equation~(\ref{QBTZ}). The geodesic is also modified. It is now described by equation
\be
z \ = \  z_h\,\frac{\sqrt{\cosh\frac{2R}{z_h}-\cosh\frac{2x}{z_h}}}{\sqrt{2}\cosh\frac{R}{z_h}} \ .
\ee
It turns out that the length of the geodesic line on figure~\ref{fig:3dT} from the vertical $z$ axis to the intersection point with curve $Q$ is given by the same expression in both $T=0$ and $T\neq 0$ cases. The total entropy is
\be
S_{\rm E} \ = \ \frac{c}{6}\log\left(\frac{1}{\pi T\epsilon}\,\sinh 2\pi TR\right) + \frac{c}{6}\log\cot\frac{\theta}{2}\,,
\ee 
as expected from the conformal transformation relating $T=0$ and $T\neq 0$ theories.  Hence the boundary entropy is again given by equation~(\ref{D2Entropy}).  We have checked that the same result for the boundary entropy can be derived in the thermal $AdS_3$ case.

\subsubsection{Constant energy density in 2+1 dimensions}
\label{sec:2+1}

It is also simple to find an example, in which the boundary entropy is not scale invariant. For example, boundary conditions discussed in the $d=2+1$ case do not preserve this property. This can be observed in the example provided by equation~(\ref{Bent2d}), in which the boundary entropy is a function of temperature. Let us study what happens with the entanglement entropy. We start from a zero-temperature case with constant surface tension boundary condition ($T_{ab}=0$, $\Sigma\neq 0$).\footnote{In this subsection we will use the results obtained earlier in~\cite{Chu:2017aab,AdSBCFTexamples} for similar configurations. Our goal will be to adapt those results to the study of the RG properties of the impurity entropy.}

Consider a single line defect~(\ref{halfspace}) in $d=2+1$ dimensions. The surface $Q$ is a plane described by equation~(\ref{Qprofile0}) with the identification between $\theta$ and $\Sigma$ provided by equation~(\ref{TTrel}). We would like to compute the change of the entanglement entropy when the defect is added to a clean system, via equation~(\ref{impentropy}). Hence we would like to study the geometric configuration similar to the one on figure~\ref{fig:3dT} (left), with an extra dimension $y$ added perpendicularly to the plane of the figure.

Before proceeding we would like to come back to the discussion of  the prescription of calculating the entanglement entropy using equation~(\ref{RTformula}) in the presence of a boundary~\cite{Chu:2017aab,AdSBCFTexamples,AdS/BCFT3}. Cutting out a slice of the bulk, as the shaded regions on figure~\ref{fig:3dT} one should be careful about the shape of the minimal surface. A new minimization problem should be considered, in which the minimal surface is allowed to end on $Q$. A simple argument shows that the minimal surface ending on $Q$ should be perpendicular to it. In other words, one needs to solve the minimization problem for the RT surface with a Dirichlet boundary condition on $M$ and a Neumann boundary condition on $Q$.

It turns out that in the $1+1$ dimensional case studied in the previous section these conditions are automatically satisfied. This is obvious in the zero temperature case, since circles and radial lines are perpendicular even in the curved $AdS_3$ geometry. One can check that the RT curves anchored on the $AdS_3$ boundary and curves $Q$ in finite temperature geometry are also perpendicular to each other. This can be seen as a simple consequence of conformal transformations, which preserve angles. This observation sheds light on the AdS/BCFT construction itself. In order to have impurity entropy independent from the CFT state, the shape of the minimal geodesics in the bulk should not depend on the impurity. The requirement is satisfied if $Q$ is determined from the Neumann boundary condition proposed by Takayanagi. 

In higher dimensional cases, the Takayanagi's boundary condition~(\ref{Neumann-metric}) with a vanishing matter stress-energy tensor $T_{ab}$ cannot guarantee that geodesics anchored on $M$ will be perpendicular to $Q$. In some sense, angles are not preserved, conformal symmetry is lost and the entropy should exhibit an RG flow.

\begin{figure}[h]
\begin{minipage}{0.45\linewidth}
\includegraphics[width=\linewidth]{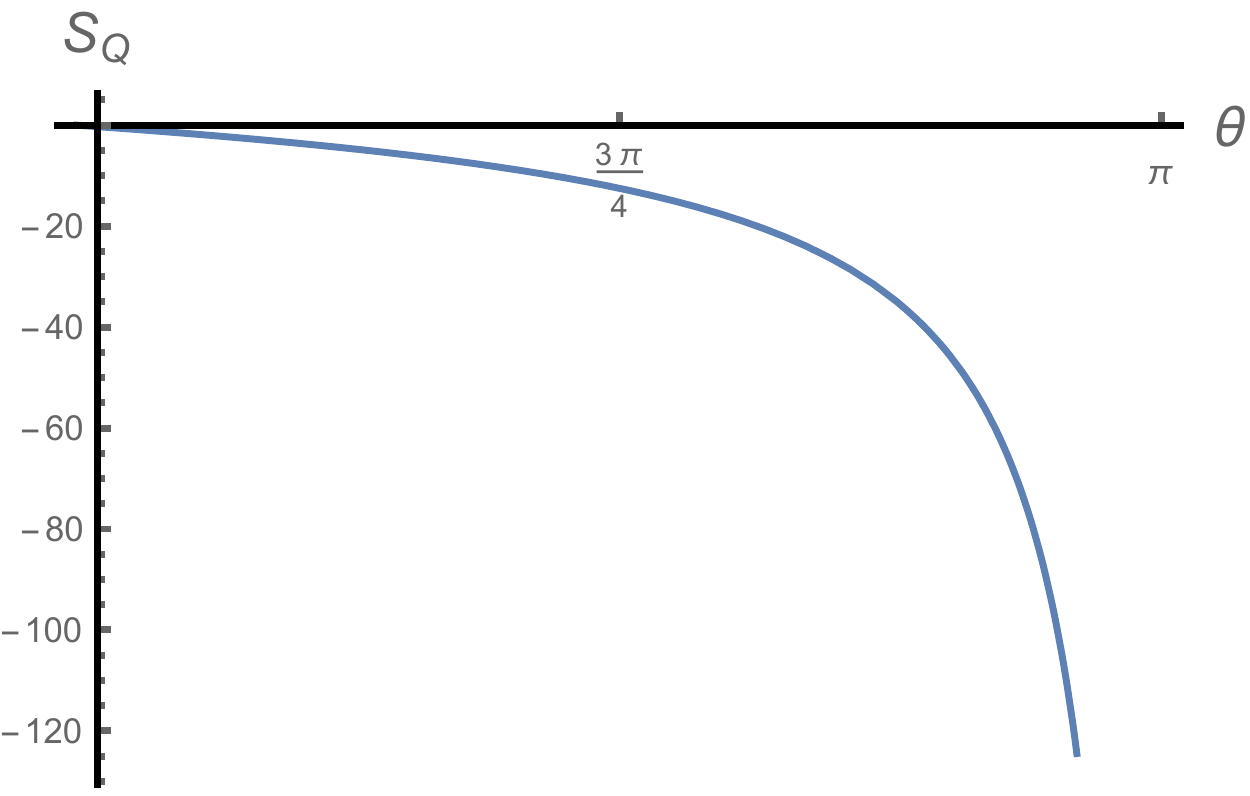}
\end{minipage}
\hfill{
\begin{minipage}{0.45\linewidth}
\includegraphics[width=\linewidth]{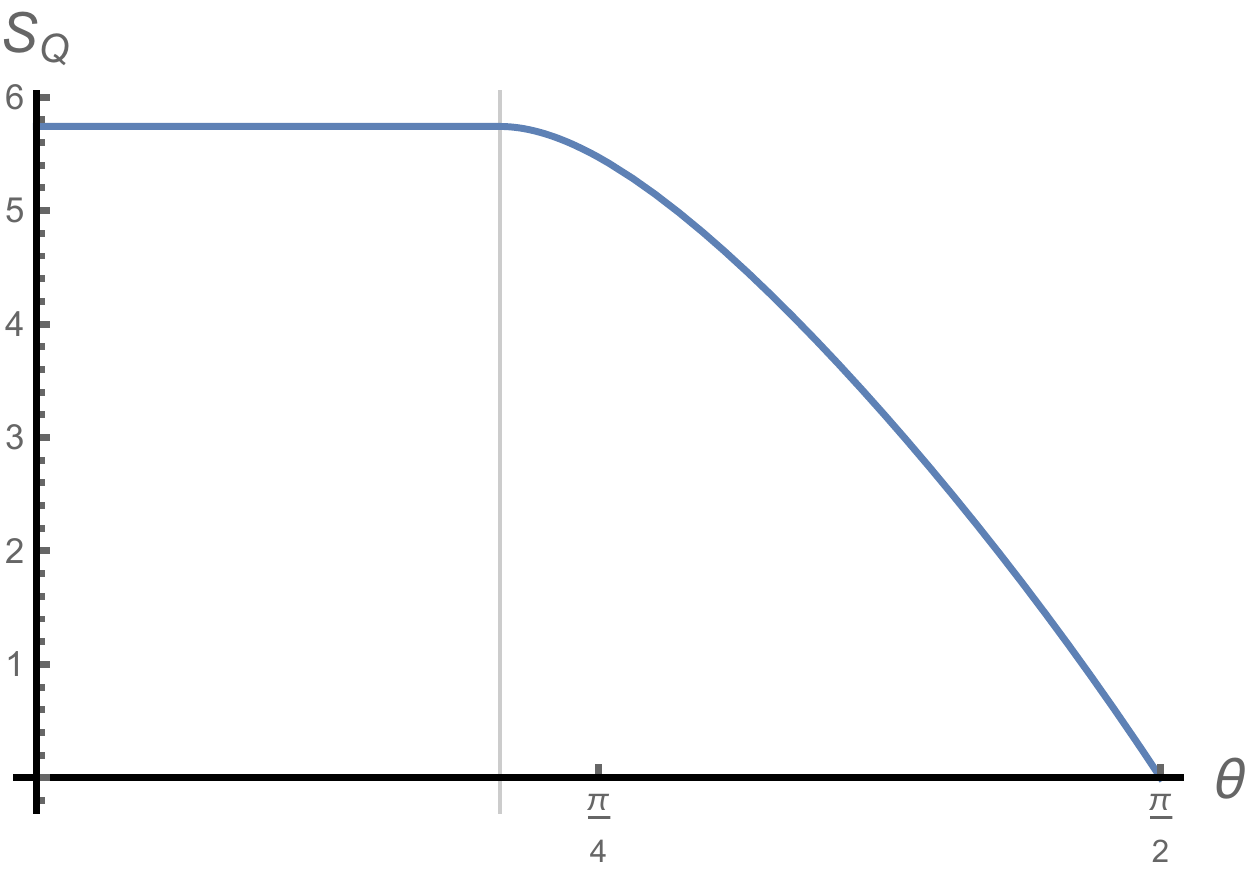}
\end{minipage}
}
\caption{Left: Entropy of a $2+1$-dimensional line defect (factor in the brackets in equation~(\ref{Simp3Dneg})) as a function of the angle for $\theta>\pi/2$. Right: Same for the case $\theta<\pi/2$.}
\label{fig:stripEnt}
\end{figure}

Without the defect the entanglement entropy is computed through the area of the minimal surface anchored on the boundary $M$ symmetrically with respect to the line $P$ of the defect. Hence, the region $A$ is chosen to be a strip of width $\Delta x$ parallel to $P$ at its center. As noted in the end of section~\ref{sec:fininterval}, such a minimal surface is defined by a profile $x(z)$, given by the incomplete beta function of equation~(\ref{beta}), translated along $y$-direction. The corresponding entropy reads~\cite{strip}:
\be
\label{SEstrip}
S_{\rm E} \ = \ \frac{L^{2}}{2G_{\rm N}}\frac{\Delta y}{\epsilon }-\frac{\pi L^{2}}{G_{\rm N}}\left(\frac{ \Delta y}{\Delta x}\right) \left[ \frac{\Gamma \left( 3/4\right) }{\Gamma \left(1/4\right) }\right] ^{2} + O(\epsilon)\,.
\ee
As before, we take a half of this expression ($R=\Delta x/2$) and subtract it from the part of the area, not cut off by the defect. In order to compute the latter piece, the intersection point of $Q$ with the minimal surface can be found following the analysis of reference~\cite{Chu:2017aab} performed in arbitrary number of dimensions. For $d=2$, the RT surface $x(z)$, anchored on one side at point $x=R$, with respect to defect $P$ placed at $x=0$, and on the other side, at some point $(x_\theta,z_\theta)$ on the surface $Q$, should satisfy the following constraints.
\be
\label{newBCs}
x(0) \ = \ R\,,\qquad x'(z_\theta) \ = \ \tan\theta\,, \qquad x(z_\theta) \ = \ - z_\theta\cot\theta\,.
\ee
Here the first two equations are the boundary conditions for the minimal surface $x(z)$, while the last condition defines parameter $z_\theta$ ($x_\theta$). Solving equation~(\ref{betaequation}) tells that the curve satisfying~(\ref{newBCs}) is
\be
\label{negbranch}
x(z) \ = \ R - \frac{z_\ast}{4} B_{\frac{z^4}{z_\ast^4}}\left(\frac{3}{4},\frac{1}{2}\right)\,, \qquad z_\ast \ = \ \frac{4R\sqrt{\sin\theta}}{\sqrt{\sin\theta}B_{\sin^2\theta}(3/4,1/2)-4\cos\theta}\,, \qquad z_\theta \ = \ z_\ast\sqrt{\sin\theta}\,.
\ee
Computing the area of the minimal surface one finds the following result for the entropy in the presence of impurity if $\pi/2\leq \theta\leq\pi$
\be
\label{Simp3Dneg}
S_Q \ = \ \frac{L^2\Delta y}{64 GR} \left[\frac{\sqrt{\sin\theta}B_{\sin^2\theta}(3/4,1/2)-4\cos\theta}{\sqrt{\sin\theta}}B_{\sin^2\theta}\left(-\frac14,\frac12\right)-B\left(\frac34,\frac12\right)B\left(-\frac14,\frac12\right)\right].
\ee
The entropy has a simple inverse dependence on the size $R$ of the entangling region. The main non-trivial part is its $\theta$-dependence, which we illustrate on figure~\ref{fig:stripEnt} (left).

The story is more interesting when $\theta<\pi/2$. In the latter case $Q$ intersects the other branch of the minimal curve satisfying
\be
\label{posbranch}
x  =  R - \frac{z_\ast}{2} B\left(\frac{3}{4},\frac{1}{2}\right) + \frac{z_\ast}{4} B_{\frac{z^4}{z_\ast^4}}\left(\frac{3}{4},\frac{1}{2}\right)\,, \quad z_\ast  =  \frac{4R\sqrt{\sin\theta}}{\sqrt{\sin\theta}\left(2B\left(\frac34,\frac12\right)-B_{\sin^2\theta}\left(\frac34,\frac12\right)\right)-4\cos\theta}\,, 
\ee
with the same relation between $z_\theta$ and $z_\ast$. While the denominator of $z_\ast$ in~(\ref{negbranch}) remains positive for $\theta>\pi/2$, the denominator of the similar expression in equation~(\ref{posbranch}) vanishes at a critical angle of $\theta_c\simeq 37^\circ$. It was observed in~\cite{Chu:2017aab} that the only available minimal surface, is the branch $x={\rm const}$, which intersects $Q$ at $z=\infty$. Consequently, at $\theta<\theta_c$ the entanglement entropy is independent from $\theta$. For $\theta_c\leq\theta\leq\pi/2$ we obtain
\be
\label{Simp3Dpos}
S_Q \ = \ \frac{L^2\Delta y}{64 GR} \left[\frac{8}{z_\ast}B\left(-\frac14,\frac12\right)-\frac{4}{z_\ast} B_{\sin^2\theta}\left(-\frac14,\frac12\right)-B\left(\frac34,\frac12\right)B\left(-\frac14,\frac12\right)\right],
\ee
where one should substitute the expression for $z_\ast$ given by equation~(\ref{posbranch}). Again, the entropy is inversely proportional to $R$. The dependence on angle $\theta$ given by the expression in the brackets is illustrated by figure~\ref{fig:stripEnt} (right). At $\theta=\theta_c$ there is a second order phase transition, so that for $\theta<\theta_c$ the entropy is independent from $\theta$. As expected the defect entropy is positive for $\theta<\pi/2$ and negative for $\theta>\pi/2$.

\subsubsection{Non-constant energy density}
\label{sec:betabrane}

We can also analyze the behavior of the entanglement entropy associated with a defect in the finite temperature geometry of section~\ref{sec:MMS}. The boundary is again a line in a $2+1$-dimensional plane and the boundary condition is defined by a $T_{ab}$~(\ref{Tab}) of a hydrodynamic form~(\ref{fluidBC}). The profile of the surface $Q$ is set by solution~(\ref{MMSprofile}). Measuring all lengths in units of $z_h$, the profile of the minimum area surface is calculated by the integral (\emph{e.g.}~\cite{Erdmenger:2017pfh}),
\be
x(z) \ = \ \pm \int  \frac{\zeta^2 d\zeta}{z_\ast ^2 \sqrt{(1-\zeta^3)(1-\frac{\zeta^4}{z_\ast ^4})}} + {\rm const}\ ,
\ee
where $z_\ast<1$ is the turning point. The analysis can be done following the steps of the previous section. In particular, the relation between the turning point and the intersection locus of the minimal surface $z_\theta=z_\ast\sqrt{\sin\theta}$ seen in solutions~(\ref{negbranch}) and~(\ref{posbranch}) also holds in the finite temperature case. However, one cannot any more solve for the turning point explicitly. For $\theta<\pi/2$ it is implicitly given by the solution of the equation 
\be
\label{turningT}
R =   2 \int\limits_{0}^{z_\ast} \ \frac{\zeta^2 d\zeta}{z_\ast ^2\sqrt{(1-\zeta^3)(1-\frac{\zeta^4}{z_\ast ^4})}} - \int\limits_0^{z_\theta} \ \frac{\zeta^2 d\zeta}{z_\ast ^2\sqrt{(1-\zeta^3)(1-\frac{\zeta^4}{z_\ast ^4})}} - \cot\theta\int\limits_0^{z_\theta}\frac{d\zeta}{\sqrt{1-\zeta^3}} \,,
\ee

As we discussed in the previous section, at zero temperature, for certain size $R$ of the region around the defect, or certain values of angle $\theta$, the minimal surface degenerates into a surface $x={\rm const}$~\cite{Chu:2017aab}. This is interpreted as a phase transition in the entanglement entropy: for $R>R_c$ it is a constant independent from $R$. Similar effect can be observed for finite temperature as well, since for sufficiently large $R$, the line $x={\rm const}$ extending from the $z=\epsilon$ to the horizon has a shorter length than the minimal curve crossing the surface $Q$. We will ignore this phase transition here, partially because of the ambiguity of the definition of the entanglement entropy at finite temperature. Following the standard approach, we will \emph{define} the entanglement entropy as the area of the surface crossing $Q$. Hence the impurity entropy is calculated by
\be
\label{SnegT1}
S_{Q} \ = \ 2\times \frac{L^2\Delta y}{4G}\left(\int\limits_{{\epsilon}}^{z_\ast}\frac{\zeta^{-2}d\zeta}{\sqrt{1-{\zeta^3}}\sqrt{1-\frac{\zeta^4}{z_\ast^4}}} + \int\limits_{{z_\theta}}^{z_{\ast}}\frac{\zeta^{-2}d\zeta}{\sqrt{1-{\zeta^3}}\sqrt{1-\frac{\zeta^4}{z_\ast^4}}} - \int\limits_{{\epsilon}}^{z_{\ast 0}}\frac{\zeta^{-2}d\zeta}{\sqrt{1-{\zeta^3}}\sqrt{1-\frac{\zeta^4}{z_{\ast0}^4}}}\right),
\ee
where the last term subtracts the entropy without the impurity, so that $z_{\ast0}$ characterizes the turning point of the minimal trajectory anchored at points $\pm R$,
\be
\label{turningT0}
R \ = \ \int\limits_0^{z_{\ast 0}}\frac{\zeta^2d\zeta}{z_{\ast 0} ^2\sqrt{1-\zeta^3}\sqrt{1-\frac{\zeta^4}{z_{\ast 0} ^4}}}\,.
\ee

\begin{figure}[h]
\begin{minipage}{0.48\linewidth}
 \includegraphics[width=\linewidth]{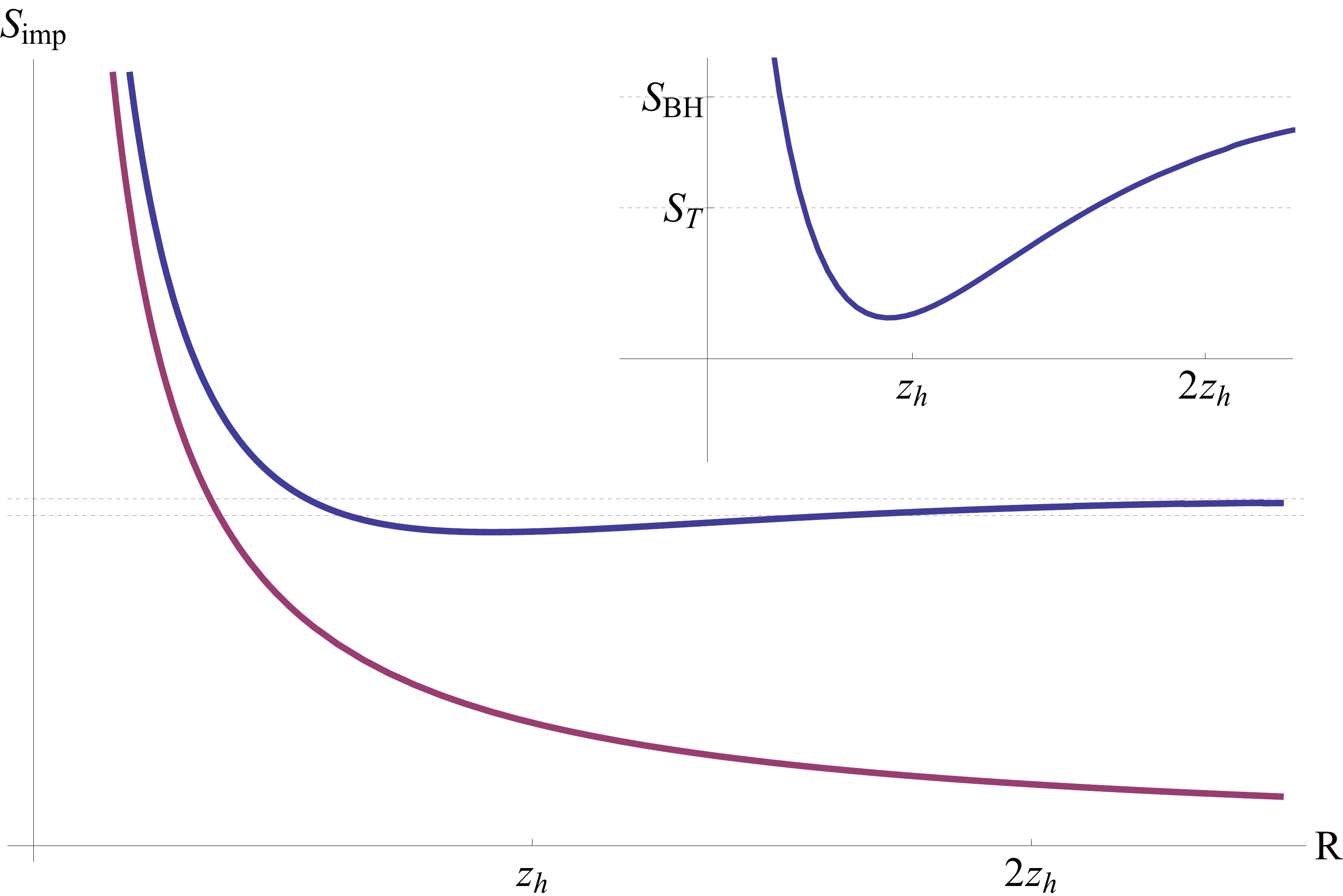}
\end{minipage}
\hfill{
\begin{minipage}{0.48\linewidth}
 \includegraphics[width=\linewidth]{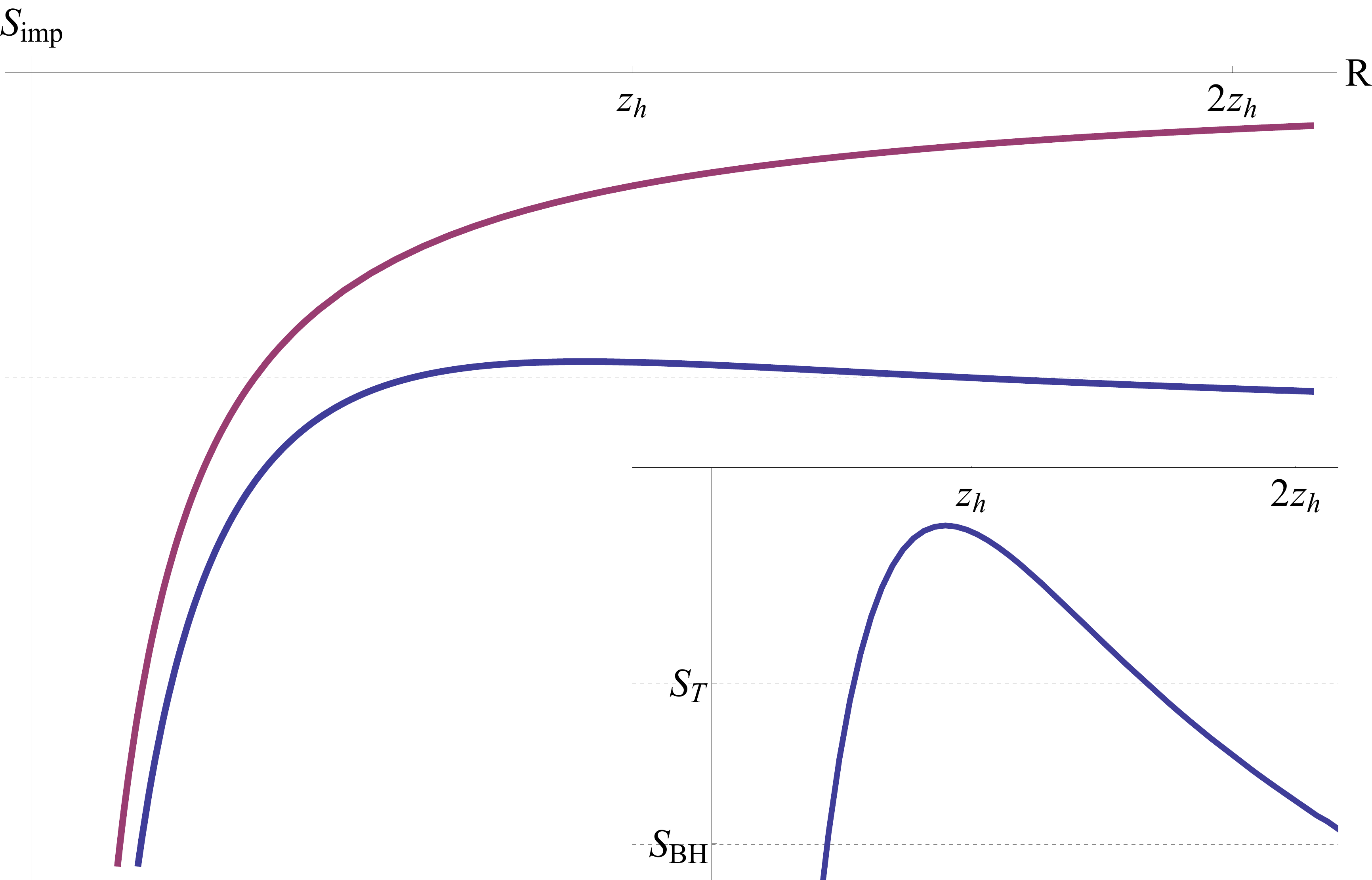}
\end{minipage}
}
\caption{Entropy of the line defect in a $2+1$-dimensional system as a function of the size of the entangling region (blue). For $R\sim 2z_h$ the result is consistent with the values predicted in~\cite{Magan:2014dwa} (inset), see section~\ref{sec:MMS}. For small regions the result asymptotes the zero temperature curve (magenta). The entropy is calculated for a ``weak'' impurity: $\theta=0.419\pi$ (left), $\theta=0.51\pi$ (right).}
\label{fig:stripEntT}
\end{figure}

The system provided by equations~(\ref{turningT})-(\ref{turningT0}) is analyzed  numerically. For the entropy to vary across a larger range of $R$ we selected the situation of a weak impurity, with $\theta\to\pi/2-0$. A characteristic form of the dependence of the entropy on $R$ is shown on figure~\ref{fig:stripEntT} (left).

For large $R$ the behavior of the entanglement entropy is consistent with both equations~(\ref{Bent2d}) and~(\ref{BHentropy}). The corresponding values are shown by the horizontal lines on the plots. Numerically the result is closer to the second definition of reference~\cite{Magan:2014dwa}, provided by equation~(\ref{BHentropy}), although we could not check this more precisely because of the difficulty of the numerical analysis for large $R$. This result is intuitively clear, because for sufficiently large $R$ the minimal curve becomes almost parallel to the horizon of the black hole. When temperature is taken to zero, one recovers the $1/R$ dependence discussed in the previous section (magenta curve on figure~\ref{fig:stripEntT}).

In case $\theta >\pi / 2$, a similar analysis produces a simpler equation relating $z_\ast$ and $R$,
\be
\label{turningT1}
R  \ =  \ \int\limits_0^{z_\theta} \ \frac{\zeta^2 d\zeta}{z_\ast ^2\sqrt{(1-\zeta^3)(1-\frac{\zeta^4}{z_\ast ^4})}}  - \cot\theta\int\limits_0^{z_\theta}\frac{d\zeta}{\sqrt{1-\zeta^3}}\,.
\ee
It can be checked from equation~(\ref{turningT1}) that the return point exceeds the radius of the horizon for some $R>R_{z_h}$. However, the intersection point $z_\theta$ is always less than $z_h$, so that the entropy is always well-defined, even for $R>R_{z_h}$. For this case, the impurity entropy is given by
\be
\label{SnegT2}
S_{Q} \ = \ 2\times \frac{L^2\Delta y}{4G}\left(\int\limits_{{\epsilon}}^{z_\theta}\frac{d\zeta}{\zeta^2\sqrt{1-\zeta^3}\sqrt{1-\frac{\zeta ^4}{z_\ast ^4}}} \ - \int\limits_{{\epsilon}}^{z_{\ast 0}}\frac{d\zeta}{\zeta^2\sqrt{1-\zeta^3}\sqrt{1-\frac{\zeta ^4}{z_{\ast 0} ^4}}}\right) \,.
\ee
The behavior of the impurity contribution to the total entropy for $\theta\to\pi/2+0$ is shown on figure~\ref{fig:stripEntT} (right) and mirrors the case $\theta>\pi/2$.

Before closing this section we would like to study the behavior of the entanglement entropy in the (1+1)-dimensional geometrical setup introduced in section~\ref{sec:fininterval}: an interval of a finite length $\ell$. That case is different from all the previous examples by the fact that boundary conditions do not have an analog of brane tension $\Sigma$, or angle $\theta$. Surface $Q$ intersects boundary $M$ at a right angle. So this is a qualitatively different type of boundary conditions, characterized by conformal $T_{ab}$ on the surface $Q$.

\begin{figure}[h]
\begin{minipage}{0.48\linewidth}
  \includegraphics[width=\linewidth]{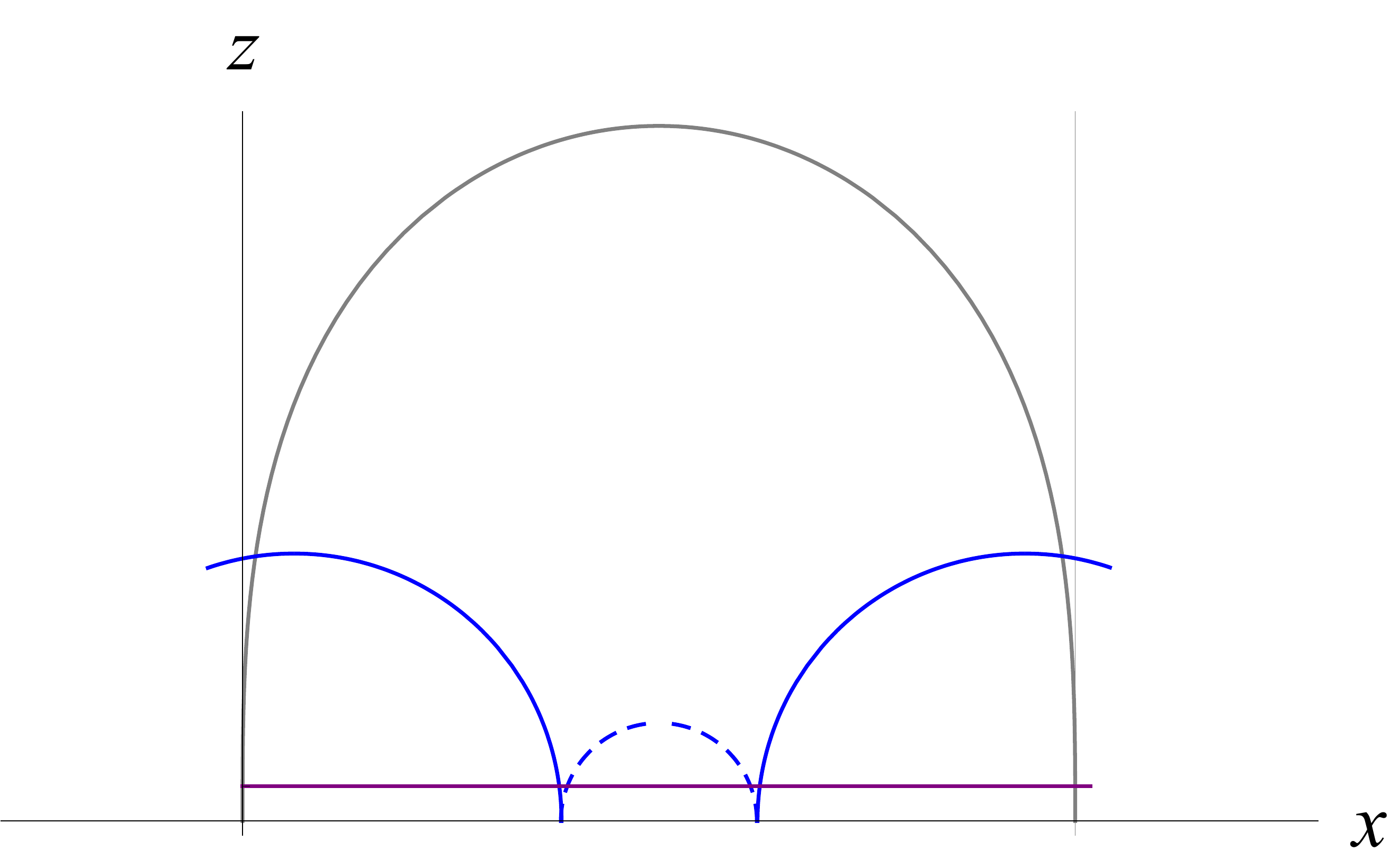}
\end{minipage}
\hfill{
\begin{minipage}{0.48\linewidth}
\includegraphics[width=\linewidth]{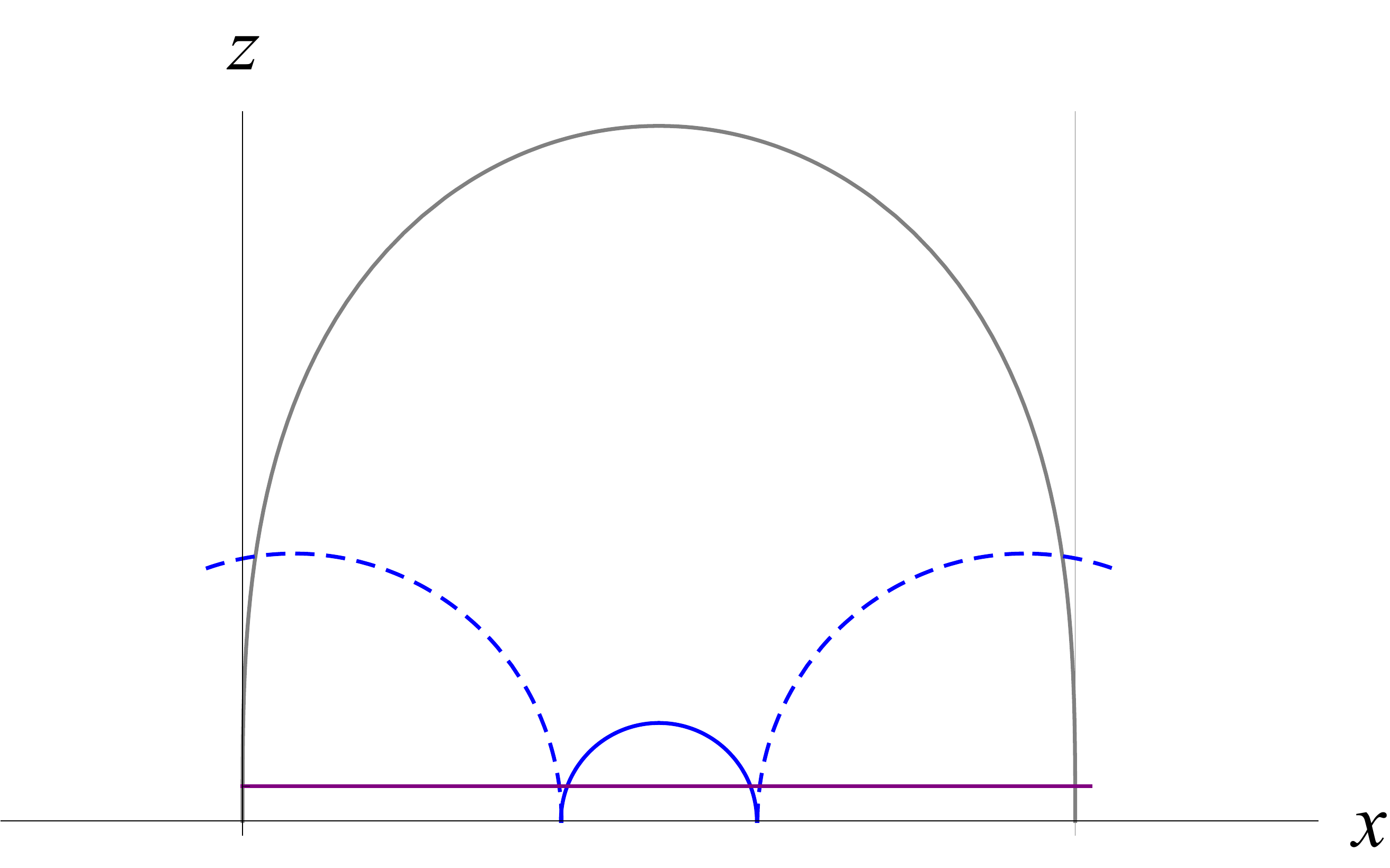}
\end{minipage}
}
\caption{Computation of entanglement entropy of two disjoint intervals in the setup with two impurities. The purple line represents the cut-off.}
\label{fig:intervals}
\end{figure}

A consequence of using different boundary conditions, in this example, is the fact that there are two impurities at two ends of a finite interval. Consequently, both impurities affect the profile of $Q$ and one should slightly modify definition~(\ref{impentropy}) in order to extract information characterizing a single impurity. Instead of computing the entanglement of a single interval around one impurity, we will assume computing the entanglement of a disjoint union of symmetric (equal length) intervals around both impurities. The resulting entropy should also be divided by a factor of two, to compute the value per impurity. This procedure is equivalent to computing the entanglement entropy, per impurity, of an infinite periodic chain of impurities on an infinite line.

\begin{figure}[h]
\begin{minipage}{0.45\linewidth}
\includegraphics[width=\linewidth]{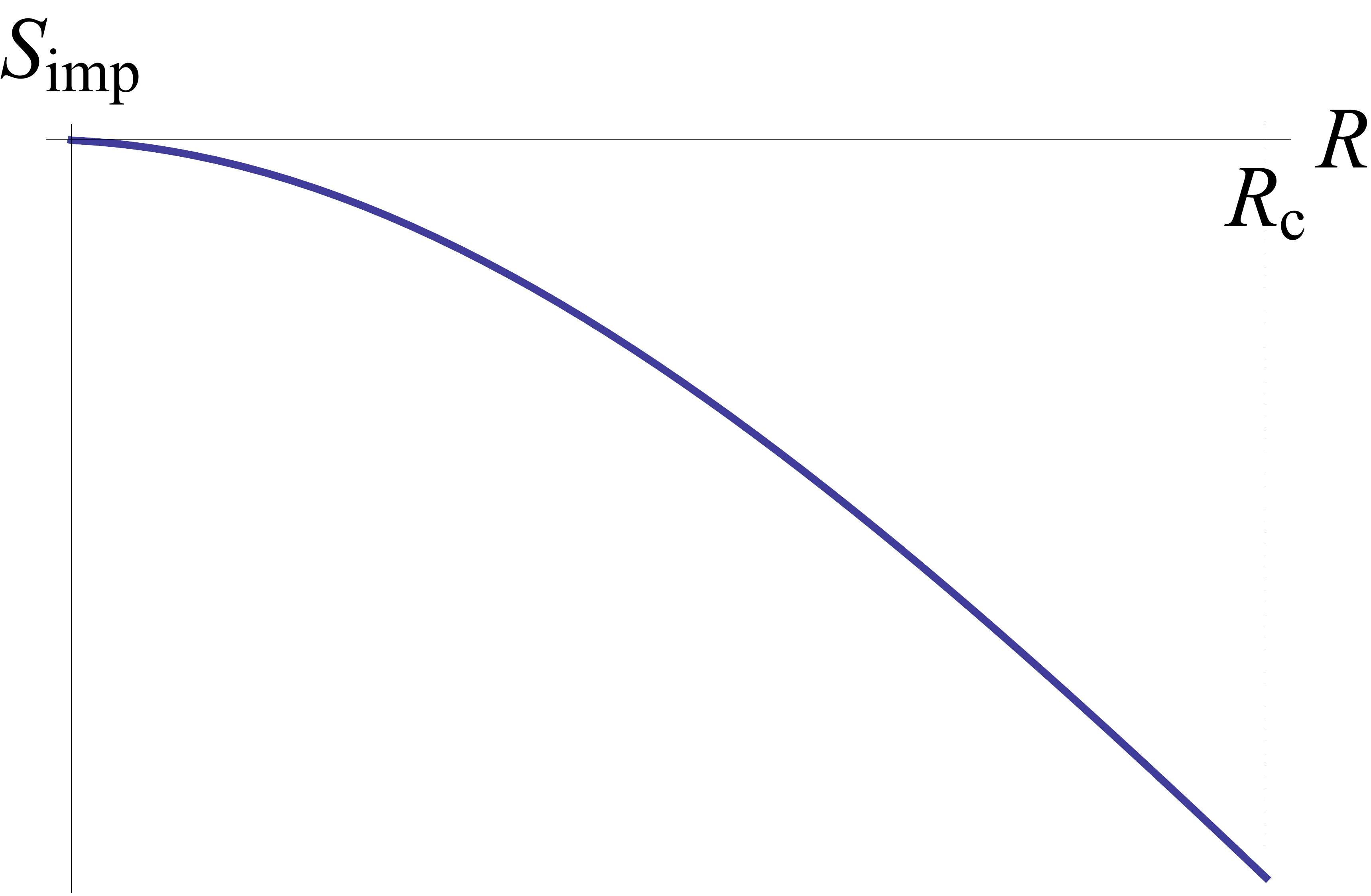}
\end{minipage}
\hfill{
\begin{minipage}{0.45\linewidth}
\includegraphics[width=\linewidth]{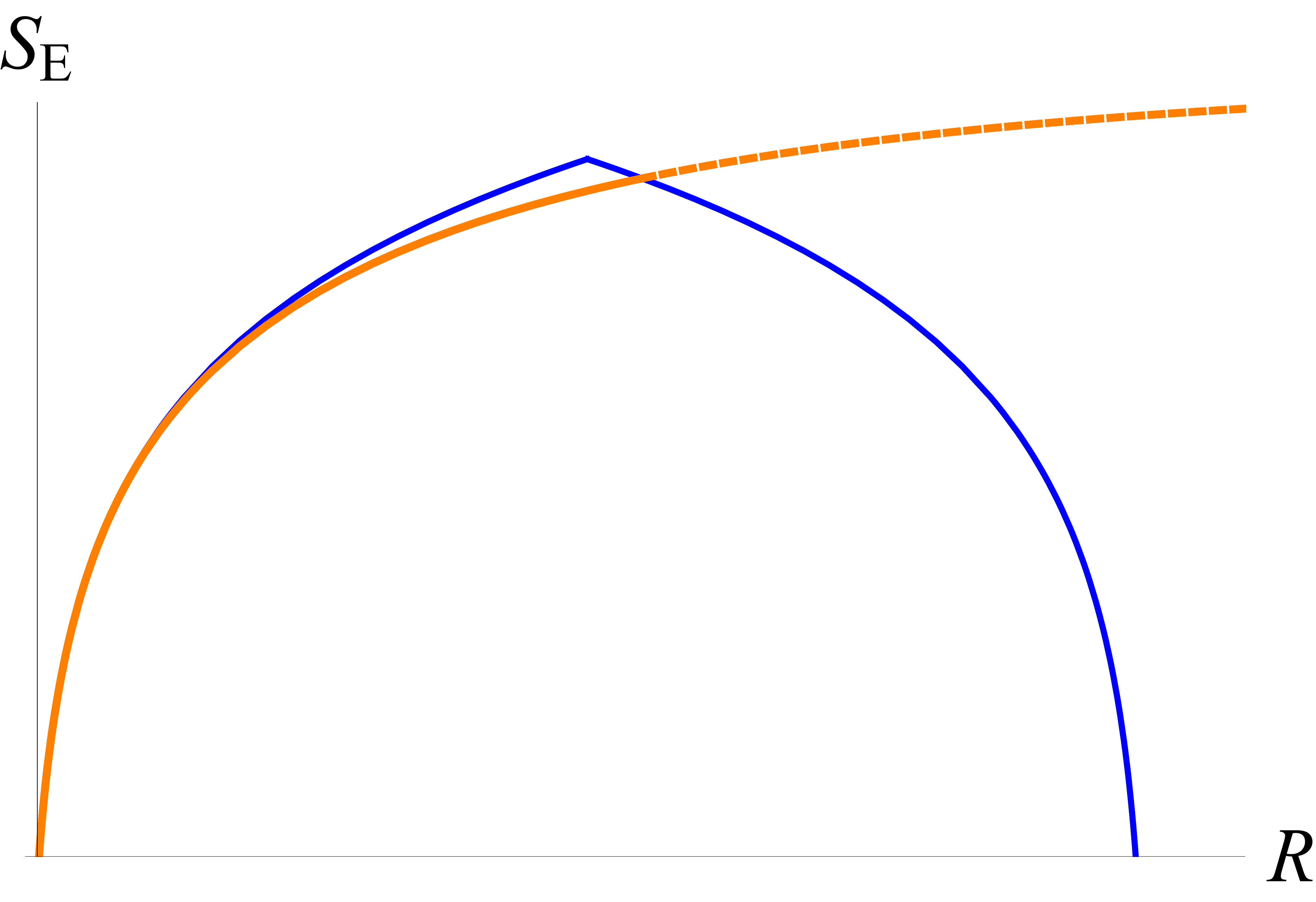}
\end{minipage}
}
\caption{(Left) Impurity entropy density as the function of size $R<R_{\rm c}$ of the entanglement interval in the system from section~\ref{sec:fininterval} for impurity separation $\ell$. (Right) Total entanglement entropy density of a periodic array of impurities (orange curve). Blue curves show the entanglement entropy of a chain of intervals in the absence of impurities, calculated per interval. Geometric transition corresponds to the cusp at the middle. In the presence of impurities the left blue curve is replaced by the the orange curve. The transition occurs at a slightly higher value of $R=R_c^\ast$, at the intersection of the orange curve with the right blue curve.}
\label{fig:BetaEnt}
\end{figure}

The geometric configuration is explained by figure~\ref{fig:intervals} (left): one has to compute the length of the segments of the arcs (solid blue) cut by the impurity curve $Q$ (gray), assuming that blue and gray lines are perpendicular. When the intervals are small the computation is essentially the same as in the previous examples, which used prescription~(\ref{impentropy}) in the setup shown on figure~\ref{fig:3dT}. In order to find the partial lengths of the geodesic lines we again use the numerical approach. Consequently, we obtain the dependence of the entropy on $R$ shown on figure~\ref{fig:BetaEnt} (left). We see that in this example the length of the part of the geodesic computing the impurity contribution grows with $R$ and the entropy is negative.

The situation is subtle when the intervals become of a size of a fraction of impurity separation $\ell$. There is a known geometric transition in the entanglement entropy of disjoint intervals in this case~\cite{Headrick:2010zt}. In the infinite chain of impurities the transition of the multipartite RT surfaces can be illustrated by the diagram
$$\begin{array}{c}
\includegraphics[width=0.3\linewidth]{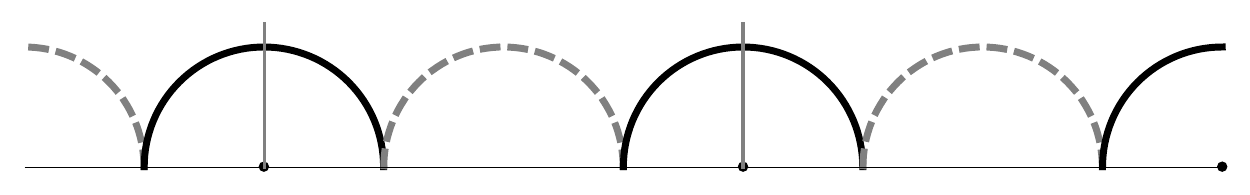}
\end{array} \qquad \longrightarrow \qquad
\begin{array}{c}
\includegraphics[width=0.3\linewidth]{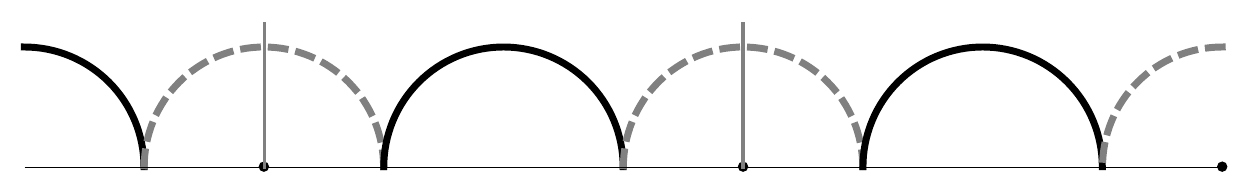}
\end{array}
$$
When the length of the geodesic lines connecting the endpoints of the intervals becomes equal to the length of the geodesics connecting endpoints of the complimentary intervals the calculation of the entanglement entropy should switch from one set of geodesics to the other.  In the clean system the transition clearly occurs when $2R_{\rm c}=\ell-2R_{\rm c}$, or $R_{\rm c}=\ell/4$. Impurities however, will introduce angle deficit (as on figure~\ref{fig:intervals}), so the transition will occur at $R>R_{\rm c}$.

After the phase transition the entropy is defined by the length of the solid arc on figure~\ref{fig:intervals} (right). Figure~\ref{fig:BetaEnt} (right) shows the behavior of the entropy of the full system in the presence and in the absence of impurities. Without impurities, the entanglement entropy of an infinite chain of intervals, computed per interval, has the behavior shown by the blue curves. Transition appears as a cusp connecting the two blue curves. Impurities lower the entropy, as illustrated by the orange curve and change the position of the phase transition (intersection of the orange and the blue curves), which then happens at $R_{\rm c}^\ast>\ell/4$. The impurities themselves decouple after the transition and do not contribute to the entropy.

\section{Discussion and Conclusions}
\label{sec:conclusions}

Let us now discuss the results obtained in the above study of holographic BCFT systems introduced by boundary conditions~(\ref{Neumann-metric}).

\subsection{Discussion of the results}

One of the goals of the present work was to test the AdS/BCFT correspondence proposed in~\cite{Takayanagi:2011zk} against some known properties of physics of boundaries and defects, in particular impurity physics in $1+1$-dimensional theories. We did it for a number of solutions reviewed in section~\ref{sec:AdSBCFT}.

In section~\ref{sec:physics} we first considered the case of an impurity in $1+1$-dimensional system whose geometric description was introduced by boundary condition~(\ref{Neumann-metric}) with $T_{ab}=0$, so that the bulk extension of the impurity was a constant-tension ``brane". We checked that such a setup is consistent with conformal boundary conditions, \emph{i.e.} BCFT. Specifically, we have checked that the impurity entropy calculated using definition~(\ref{impentropy}) is consistent with the BCFT expectations. The entanglement entropy contribution of the impurity, determined in section~\ref{sec:TEntropy}, is independent from the temperature and from the size of the entanglement region -- it only depends on the boundary condition itself (tension of the brane). This entropy, expressed by equation~(\ref{SQ}), is also equivalent to the thermodynamic entropy of the impurity computed in section~\ref{sec:1+1}. While on the CFT side it is a consequence of conformal symmetry, on the gravity side it might seem like a non-trivial geometric fact. Although the calculations reported in this paper are performed on a more general class of geometries they essentially confirm the earlier results of paper~\cite{Takayanagi:2011zk}.

We also note that in the 2+1 dimensional geometry the geodesics anchored on the boundary are automatically orthogonal to the profiles of surfaces $Q$ satisfying~(\ref{Neumann-metric}) with $T_{ab}=0$. In a way this explains the meaning of the Neumann boundary conditions imposed by Takayanagi. They fix the angle at which the geodesic lines extended from the boundary intersect $Q$. Due to the properties of the conformal transformations this guarantees that the impurity entropy is state independent.

Impurity entropy in equation~(\ref{SQ}) is expressed in terms of the geometric parameter -- angle $\theta$, at which the boundary brane $Q$ is intersecting the boundary of AdS space (as on figure~\ref{fig:impent}). When $\theta<\pi/2$, entropy is a positive number. However, for $\theta>\pi/2$ entropy is negative. There is no contradiction, since this entropy is introduced as a difference of the entanglement entropy with impurity present and the one with no impurity present. Consequently, negative result means that impurity reduces the number of degrees of freedom. This happens when the interaction of the impurity with the system, is attractive. For example, the impurity can form bound states. Consequently, the repulsive interaction is characterized by a positive relative entropy.

The two situations discussed in the previous paragraph can refer to different signs of the coupling in equation~(\ref{CFTdeform}). Indeed, understanding of the correspondence between the geometric construction and the CFT deformation operator was another motivation of the present paper. Since the entropy is a scale independent constant one can conclude that this deformation  is marginal. Below we will discuss examples of relevant and irrelevant deformations.

It is known in general that an impurity can have a drastic effect on entanglement of two sides of a one-dimensional system separated by it. Consider a 1D conformal system of length $2R$ with an impurity in the middle. The entanglement entropy of either half of the system should be
\be
\label{halfspaceent}
S_{\rm E} \ = \ \frac{c}{6}\,\log\frac{R}{\epsilon} + O(\epsilon^0)\,,
\ee
if no impurity is present. It was argued in~\cite{BulkImp} and confirmed by numerical experiments that if impurity corresponds to a relevant perturbation of the CFT, then it effectively disentangles the two sides of the system, \emph{i.e.} the resulting entanglement entropy, when the relevant perturbation is turned on, drives $S_{\rm E}$ to zero.

This effect can, for example, be studied in a XXZ Heisenberg chain~\cite{Affleck:review}. The Hamiltonian of the system is given by
\be
H_{XXZ} \ = \ \sum J_j\left(S_j^xS_{j+1}^x+S_j^yS_{j+1}^y + \Delta S_j^z S_{j+1}^z\right)\,.
\ee
This Hamiltonian describes a conformal fluid in the large wavelength limit if $-1\leq \Delta\leq 1$. Coupling $J$ is unity for all links except the middle, impurity link, which has $J_{\rm imp}<1$. Such a situation describes a relevant perturbation of the CFT if $\Delta>0$ (repulsive interaction) and irrelevant one if $\Delta<0$ (attractive).

The RG flow of the chain is studied by computing the entropy as a function of the size $R$. It was shown that with $R$ increasing the ``effective central charge'' (the coefficient of the $\log(R/\epsilon)$ term) renormalizes to the original value $c/6$ in the attractive case ($\Delta<0$) and to zero in the repulsive case ($\Delta>0$).

It is straightforward to geometrically illustrate the fixed points of such an RG flow. For this we will think about them in terms of the compact space. (Appendix~\ref{global} explains the AdS/BCFT construction in global compact coordinates in $AdS_3$.) In global coordinates, the constant-tension brane connects opposite points of the spatial circle (as in figure~\ref{fig:gads}). If tension $\Sigma$ is positive (angle $\theta<\pi/2$) then the center of the AdS space falls in the interior of the bulk region $N$, which encodes boundary system $M$. If tension is negative ($\theta>\pi/2$) $N$ is less than a half of the AdS space. (We remind that $M$, $N$ and $\theta$ are defined by figure~\ref{fig:NMQP}, while equation~(\ref{TTrel}) relates $\Sigma$ and $\theta$.)

\begin{figure}[h]
  \centering
  \includegraphics[width=0.4\linewidth]{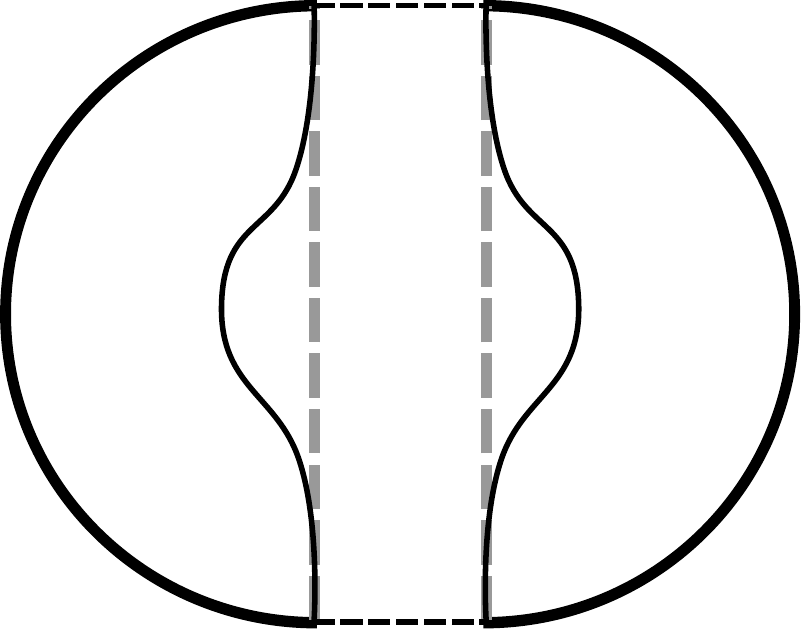}
  \caption{Two parts of $AdS_3$ glued along the $Q$-brane with negative tension. The gray dashed line is the minimal RT surface connecting two ends of the halfspaces on the boundary.}
  \label{fig:bulkimp}
\end{figure}

In order to describe impurities, which separate two intervals we glue together two spaces $N$ obtained from two copies of $AdS_3$ space cut along the boundary $Q$ as shown on figure~\ref{fig:bulkimp} in the case of negative tension. In such a case we are dealing with two impurities separating a circle into two equal arcs.

To determine the entanglement entropy of each arc one needs to compare the configuration of boundary $Q$ with that of the minimal surface (geodesic line). It is obvious that for positive tension the minimal surface lies inside $N$, while for the negative tension it belongs to the exterior. Consequently, in the first case, the entanglement entropy of each arc is given by equation~(\ref{halfspaceent}) with an extra factor of two for two impurities. In the second case the entanglement entropy is zero. Therefore, the geometric cartoon indeed illustrates the properties of the two fixed points, with $\Delta$ and $\Sigma L=\cos\theta$ playing a similar role.

A $d=2+1$-dimensional example of a constant-tension brane, considered in section~\ref{sec:2+1} gives a geometric example of an RG flow. The entropy computed by equations~(\ref{Simp3Dneg}) and~(\ref{Simp3Dpos}) is inversely proportional to the size $R$ of the strip around the impurity. For both $\theta<\pi/2$ and $\theta>\pi/2$ the absolute value of the impurity entropy is a decreasing function of $R$, which asymptotes to zero. Hence in those two cases one has a repulsive or attractive irrelevant deformation.

In the $d=2+1$ dimensional finite temperature example considered in section~\ref{sec:betabrane}, one can study the RG flow by either considering temperature, or $R$-dependence of the entropy. In both cases the result is consistent with the zero temperature analysis, showing that the deformation is irrelevant. The finite temperature behavior connects to the results of reference~\cite{Magan:2014dwa}, where two inequivalent definitions of the defect entropy were considered. The present work indicates that entanglement entropy at finite temperature is consistent with the definition of the defect entropy as the Bekenstein-Hawking coefficient times the area of the horizon shadow of the bulk extension of the defect. We also observe a non-monotonic renormalization of the entanglement entropy to this value (see figure~\ref{fig:stripEntT}).

An interesting example of boundary systems is provided by solutions with $Q$ defined by a conformal stress-energy tensor $T_{ab}$ in equations~(\ref{Neumann-metric}). Such solutions were considered in section~\ref{sec:fininterval}, assuming $\Sigma=0$. In particular, the corresponding profiles of bounding curve $Q$ intersect boundary $M$ at a right angle and have nice scaling properties. Moreover $Q$ in this case parameterizes a minimal area surface bounded by a strip in a dimension larger by one. This observation gives an implicit relation between the entanglement entropy of a strip of width $\ell$ in $d+1$ dimensions and a BCFT on a strip with the same width in $d$ dimensions subject to this special boundary condition.

The presence of two impurities at the boundaries of a finite system $M$ affects the RG behavior of the entropy. This boundary condition corresponds to a relevant deformation (figure~\ref{fig:BetaEnt}), since the impurity entropy grows as the energy scale is decreased. The dependence of the entanglement entropy of the interval starting at an impurity, as a function of the interval size, shows a characteristic cusp due to the geometric transition, and an asymmetry due to the impurities, which can, in principle, be compared with numerical DMRG simulations in spin chains.

The transition itself is characterized by a decoupling of the impurity. For the interval sizes larger than the critical scale, the contribution of the impurity to the entropy vanishes. The critical scale can then be interpreted as a screening radius of the impurity.

\subsection{Outlook}

We now briefly mention some open questions and possible future research directions. First of all, we are not aware of the exact dictionary translating the parameters of the holographic AdS/BCFT construction to the language of the dual field theory. In particular, it would be interesting to define $\theta$ in purely CFT terms. Naturally, we expect that equation~(\ref{BEntropy}) should be useful in this context. We would also like to better understand the physical relevance and the dual interpretation of the class of solutions of~\cite{Erdmenger:2014xya}, reviewed and generalized in section~\ref{sec:fininterval}. Let us further discuss their properties. 

We have already mentioned that the profiles of the bulk bounding surface $Q$ implicitly relate the AdS/BCFT problem with the calculation of the entanglement entropy in a higher dimensional setup. For example, given an interval of length $\ell$ in $1+1$ dimensions, the profile $x(z)$ of $Q$ can be found from solution~(\ref{beta}) after an appropriate rescaling. Now the entanglement entropy of an infinite strip of width $\ell$ in a $2+1$-dimensional case can be computed using the same $x(z)$:
\be
S_{E} \ = \ \frac{2\Delta y}{4G} \int\limits_{\epsilon}^{z_\ast(\ell)} dz\ \frac{L}{z} \sqrt{\frac{L^2}{z^2} + \frac{L^2}{z^2}x'^2(z)}\,.
\ee
This relation is generalizable to higher dimensions. It would be interesting to understand if this holographic relation can be quantified in terms of dual quantities, like entanglement entropies. Indeed, it is known that in some cases the $d=1+1$ CFT impurity entropy, is related to the universal topological entanglement entropy of certain $d=2+1$ configurations~\cite{Kitaev:2005dm}.

A geometric quantity, which characterizes solutions considered in section~(\ref{sec:fininterval}), is the length of the bounding profile $Q$. A straightforward computation shows, that this length is given by
\be
 2L\int\limits_\epsilon^{z_\ast} \frac{dz}{z\sqrt{1-z^4}} \ = \   L\log \left(\frac{z^2}{1+\sqrt{1-z^4}}\right)\bigg|_{\epsilon}^{z_\ast}\,,
\ee
where a cutoff was introduced in the same way as in the calculation of the RT entropy. The upper integration limit gives zero contribution and the lower yields the following result
\be
\int dt_1 \int dt_2 \ \langle {\cal O}(x_1,t_1){\cal O}(x_2,t_2)\rangle \ = \ \frac{c}{6}\,\log\frac{|x_1-x_2|}{\epsilon} + \frac{c}{12}\log 2 + O(\epsilon)\,,
\ee
where we expressed the result in terms of some correlation function by dividing the length by $4G$. We see that the leading divergence of the length of $Q$ is the same as of the entanglement entropy. Indeed, the leading divergence is completely determined by the asymptotic part of the curve, close to the AdS boundary. There it coincides with the RT minimal curve.

Some other directions were left beyond the scope of the present paper. In appendix~\ref{appendix} we have reviewed other instances of the AdS/BCFT problem. In particular, in sections~\ref{sec:disc} and~\ref{sec:newsols} we discussed time-dependent solutions. We believe that such solutions may be interesting in the study of non-equilibrium problems in the holographic approach. It would also be interesting to look for AdS/BCFT solutions with mixed boundary conditions. Such solutions are also interesting in the CFT context. We leave further comments on these and other questions raised in this section for a future work.

\paragraph{Acknowledgements} We would like to thank Carlos Hoyos, Horatiu Nastase, Fabio Novaes and Ara Sedrakyan for the stimulating discussions. The work of all the authors was supported, at least partially, by the Brazilian Ministry of Education. The work of DM was also partially supported by the joint Armenian-Russian collaboration grant of the Russian Foundation of the Basic Research, RFBR-18-51-05015\_Arm. DM would also like to thank FAPESP grant 2016/01343-7 for funding his visit to ICTP-SAIFR in August 2018 as well as the Nordita Institute in Stockholm for supporting the visit in June-July 2018,  where a part of this work was done. The work of MS was partially supported through the Brazilian Professor fellowship program of the International Institute of Physics in Natal.

\begin{appendix}

\section{More examples of the AdS/BCFT construction}
\label{appendix}

\subsection{Disk boundary}
\label{sec:disc}

A number of new solutions to AdS/BCFT boundary conditions~\ref{Neumann-metric} can be generated by applying isometries on basic solution~(\ref{Qprofile0}) in section~\ref{sec:basic}. The $AdS_{d+1}$ metric is invariant under the $d$-dimensional conformal group, which contains translations, rotations, boosts, dilatations and special conformal transformations. As a first non-trivial example we consider a special conformal transformation following the original results of references~\cite{Takayanagi:2011zk,AdS/BCFT2}.

In the Euclidean space, $t\to iy$, the half-plane $x>0$ on the boundary can be  mapped to the interior of a disc by a global transformation
\be
\label{SCT}
x'_\mu \ = \ \frac{x_\mu + c_\mu x^2}{1+2(c\cdot x)+c^2x^2} \,,
\ee
where $c_\mu$ is a constant vector and $x^\mu=(x,y)$. The map of the half-plane $x=0$ to the disc of radius $R$ corresponds to the choice $c_\mu = (1/2R,0)$. The transformation maps lines of constant $x>0$ to circles of radii $r<R$. Infinity is mapped to the point $(2R,0)$. The AdS metric is invariant under this transformation if the coordinate $z$ is transformed according to
\be
z' \ = \ \frac{z}{1+2(c\cdot x)+c^2x^2}\,.
\ee
In the bulk the transformation maps the two-dimensional Euclidean $AdS_2$ slices, including the hypersurface $Q$ of section~\ref{sec:basic} into spherical  domes ending on $M$. The new $Q$ is then defined by equation
\be
\label{QSphere}
y^2 + \left(x - R\right)^2 + \left(z - z_0\right)^2 \ = \ R^2 \csc ^2 \theta\,, \qquad z_0 \ = \ - R\cot\theta\,.
\ee
As before, $\theta$ is the external angle of intersection of the spherical surface with the $z=0$ boundary. When tension $\Sigma=0$, $\theta=\pi/2$, $Q$ is exactly a hemisphere. In the case of positive tension  ($\theta<\pi/2$), the center of the sphere is at $z_0>0$ and \emph{vice versa}.

It is also interesting to consider the analytic continuation of the spherical solution to the Minkowski space.
\be
\label{Qhyper}
-t^2 + \left(x - R\right)^2 + \left(z - z_0\right)^2 \ = \ R^2 \csc ^2 \theta\,.
\ee
The new solution describes a compact space, whose walls are expanding. Gluing the Euclidean and Minkowskian solutions at $t=y=0$, one obtains a solution of the bubble creation problem, where a bubble of the size $R$ is created at $t=0$ and  expands for $t>0$. One typically thinks of such a solution in the context of a phase transition, when a bubble of true vacuum is created inside a false one. The true vacuum is represented here by the anti-de Sitter space, while the effect of the false vacuum is introduced effectively through the non-zero surface tension.

\begin{figure}[htb]
\begin{minipage}{0.45\linewidth}
 \includegraphics[width=\linewidth]{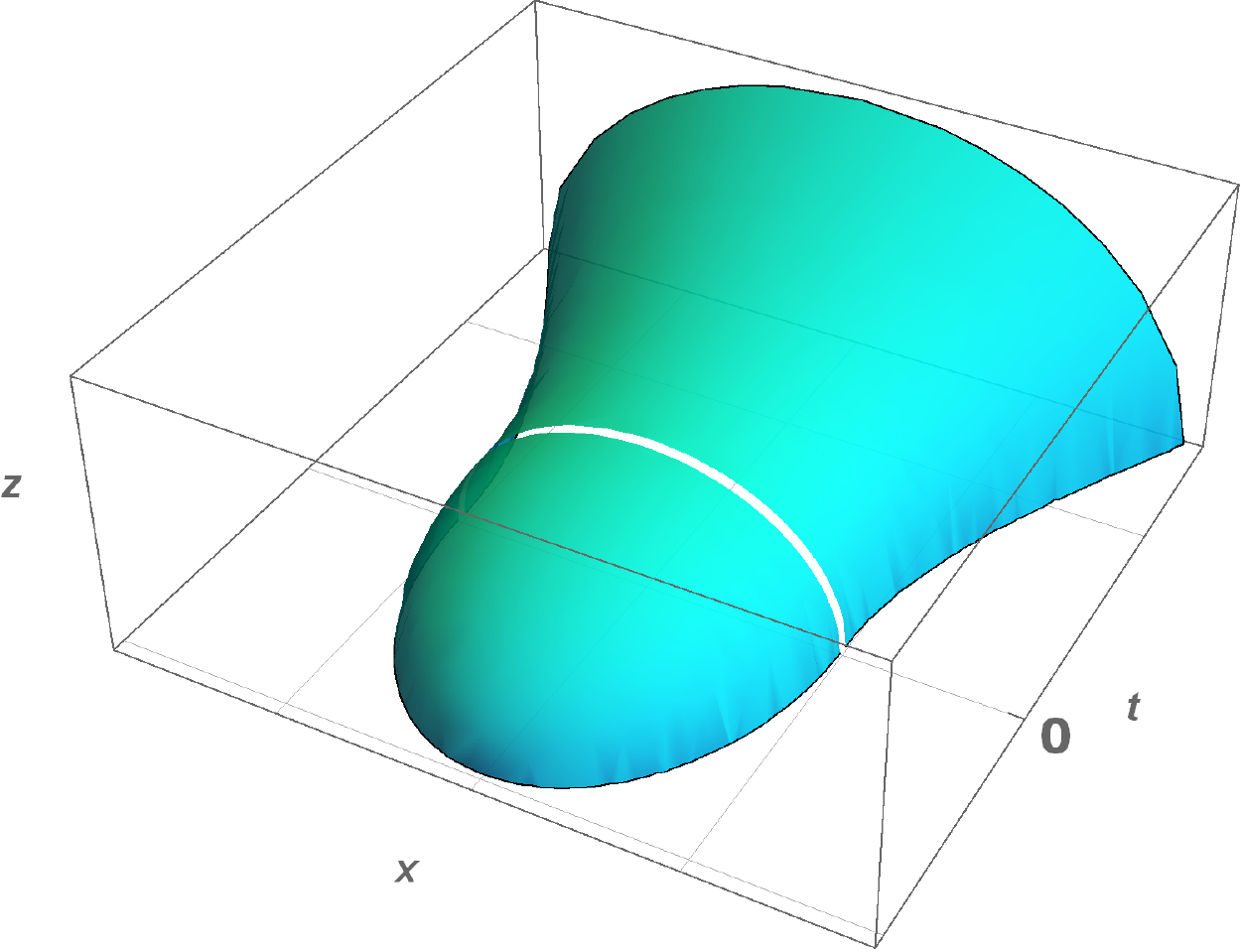}
\end{minipage}
\hfill{
\begin{minipage}{0.45\linewidth}
 \includegraphics[width=\linewidth]{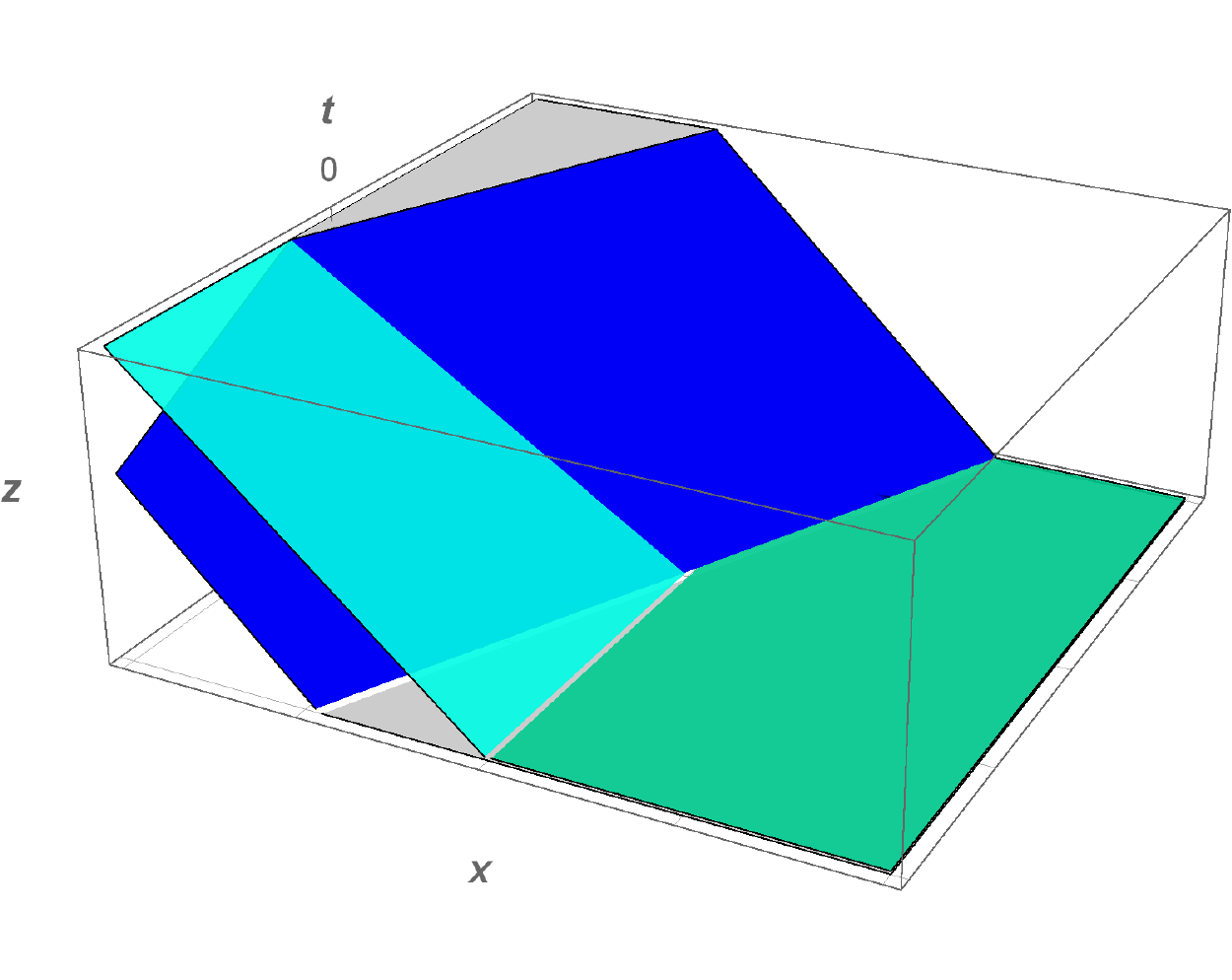}
\end{minipage}
}
 \caption{(Left:) The nucleation of a Euclidean bubble of anti-de Sitter space ($t<0$), creation ($t=0$) and evolution of the real bubble ($t>0$). (Right:) Stationary (cyan) versus boosted (blue) interfaces.}
 \label{fig:bubble}
\end{figure}

Solution~\ref{QSphere} can be extended to arbitrary dimension~\cite{AdS/BCFT2}. $AdS_3$ case is illustrated on figure~\ref{fig:bubble}. Similar solutions has been recently discussed in the context of the black hole escapability problem in~\cite{Almheiri:2018ijj}.

\subsection{Boosted boundary}
\label{sec:newsols}

The second non-trivial transformation, which can be applied to solution~(\ref{Qprofile0}) is a boost. For boost rapidity $\eta$ one finds
\be
\label{Qboosted}
x(t,z) \ = \  \left(\tanh\eta\right)t + \left(\frac{\cot\theta}{\cosh\eta}\right) z\,,
\ee
Apart from a hyperbolic rotation of $Q$ in the $x-t$ plane, the boost changes the apparent angle in the $x-z$ plane at which the hypersurface intersects the boundary $z=0$. The actual angle remains $\theta$.

This solution can be compared with solution~(\ref{Qhyper}). Equation~(\ref{Qboosted}) describes an interface moving with a constant velocity. In the $z$-direction the interface is tilted, such that the angle depends on the velocity. Profile~(\ref{Qhyper}) is a dynamical solution describing a finite size interface characterized by an additional scale $R$, which undergoes an accelerated expansion. The walls of the bubble~(\ref{Qhyper}) move asymptotically ($t\to\infty$) with a speed of light. Hence, at early times (small velocity) the tilt angle in the $x-z$ plane is close to $\theta$. At late times the angle asymptotes $\pi/2$ as
\be
{\rm arccot}\left(\frac{R}{t}\cot\theta\right)\,.
\ee
Thus, at small $z$ the two solutions are similar in the sense that at any given $t$ the bubble walls move as an interface of type~(\ref{Qboosted}) with $\cosh\eta \ = \ t/R$.

\section{AdS/BCFT solutions in global coordinates}
\label{global}

It is useful to reobtain some of the above results in global  anti-de Sitter space. The solutions of equations~(\ref{Neumann-metric}) in the global setup were perhaps originally discussed in~\cite{Azeyanagi:2007qj} and later in~\cite{Erdmenger:2014xya}. More recently similar solutions reappeared in~\cite{Almheiri:2018ijj}.

Let us work in the metric
\be
\label{globalBTZ}
\frac{ds^2}{L^2} \ = \ d\rho^2 - \frac{1}{4}({\rm e}^{\rho}+J{\rm e}^{-\rho})^2dt^2 +\frac{1}{4} ({\rm e}^{\rho}-J{\rm e}^{-\rho})^2d\phi^2\,.
\ee
Here $\rho$ is the ``holographic'' coordinate, with the conformal boundary at $\rho\to\infty$, and $0\leq\phi<2\pi$ is a compact boundary spatial coordinate. For $J=1$ this solution is an empty $AdS_3$ space. For $J <0$ this solution describes a BTZ black holes with temperature
\be
T_H \ = \ \frac{\sqrt{-J}}{2\pi}\,.
\ee
For the remaining positive values of $J$ this metric has a naked (conical) singularity at the origin, $\rho=0$. Such solutions would be analogs of the thermal AdS solutions in the Poincar\'e coordinates, considered in section~\ref{sec:TAdS}, albeit with no condition for the spatial cycle to be contractible in the bulk.

It is also convenient to work with an explicitly compact set of spatial coordinates, so that the metric of the black hole takes the form
\be
\label{globalcompact}
\frac{ds^2}{L^2} \ = \ \frac{\alpha^2}{\cos^2\chi}\left(\frac{d\chi^2 }{\alpha^2}- \sin^2\chi\, dt^2 + d\phi^2 \right)\,,
\ee
with $0\leq \chi \leq \pi/2$ being a new holographic coordinate, the boundary of AdS space at $\chi=\pi/2$ and Hawking temperature $T_H\ = \ \alpha/2\pi$.

\subsection{AdS/BCFT in empty global \texorpdfstring{$AdS_3$}{AdS3}}

\begin{figure}[htb]
 \centering
 \begin{minipage}{0.4\linewidth}
  \includegraphics[width=\linewidth]{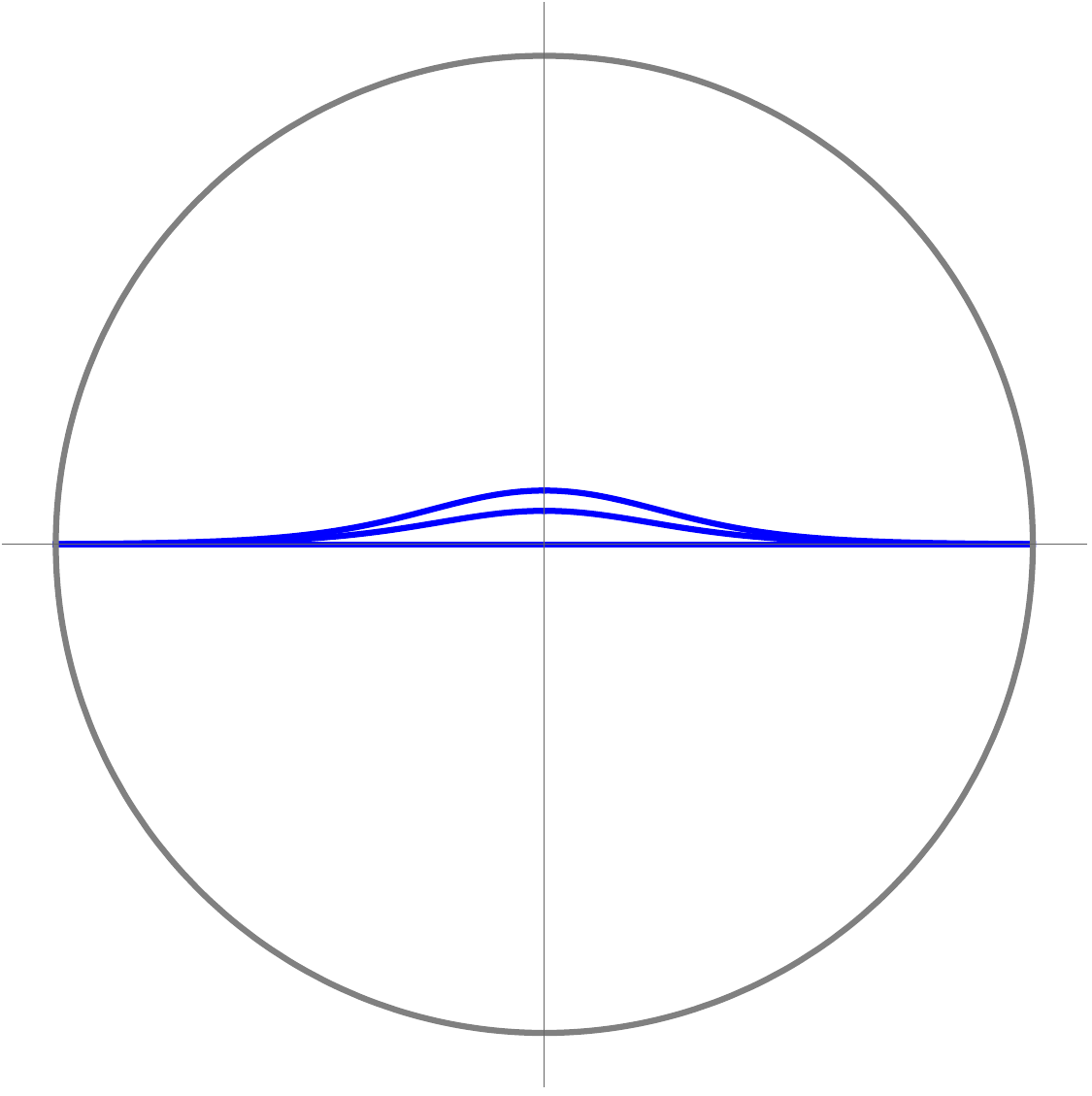}
 \end{minipage}
\hfill{
\begin{minipage}{0.4\linewidth}
  \includegraphics[width=\linewidth]{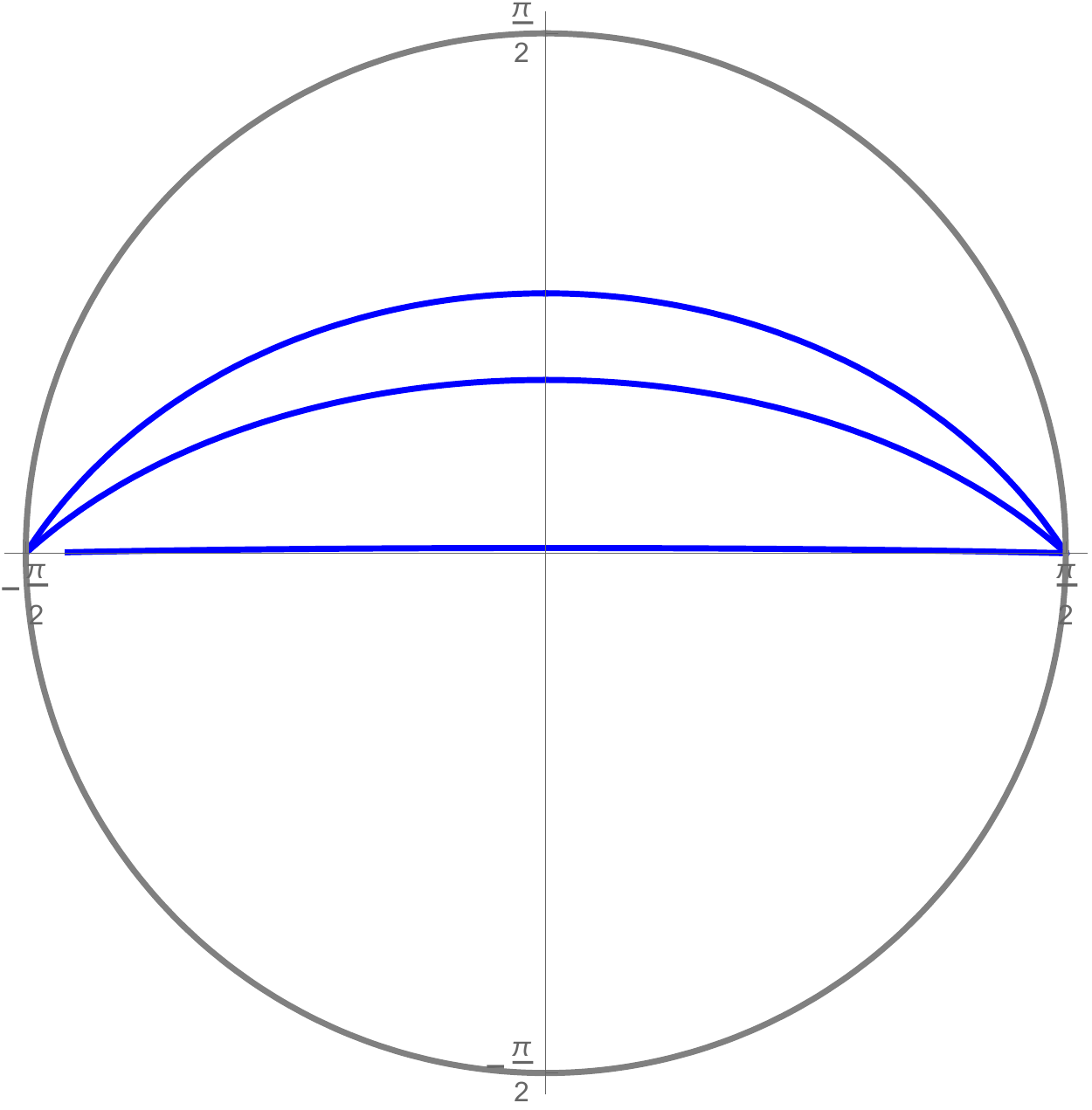}
 \end{minipage}
}
 \caption{Profiles of $Q$ in empty global $AdS_3$ (fixed time slice) in non-compact (left) and compact (right) coordinates for $\theta=\pi/2,\pi/3,\pi/4$.}
 \label{fig:gads}
\end{figure}

We parameterize the interface $Q$ as $\phi=\phi(\rho)$. For $J=1$ there is an analytical solution to equation
\be
K_{ab} - (K-\Sigma)h_{ab} \ = \ 0\,.
\ee
With a boundary condition $\phi(\infty)\ = 0$ the solution is
\be
\phi =  \arccot\left(\sqrt{\sinh^2\!\rho\,\tan^2\theta-1}\right).
\ee
As before we introduce $\cos\theta=\Sigma L$. In the global picture the case $\Sigma=0$, or $\theta=\pi/2$, corresponds to the profile of $Q$ cutting the AdS space in two halves along the diameter. For other values of $\theta$, $Q$ crosses the bulk avoiding the center by the maximal approach at $\rho=\rho_\ast$,
\be
\sinh\rho_\ast \ = \ {|\cot\theta|}\,.
\ee
Several profiles of $Q$ are demonstrated on figure~\ref{fig:gads}(left).

In terms of compact coordinates~(\ref{globalcompact}) the solution is
\be
\phi \ = \ \arctan\left(\frac{\cos\theta\cos\chi}{\sqrt{\sin^2\theta-\cos^2\chi}}\right)\,.
\ee
The turning point corresponds to $\chi=\chi_\ast$ with $\cos\chi_\ast=\sin\theta$.

\subsection{AdS/BCFT in the global BTZ geometry}

Using compact coordinates~(\ref{globalcompact}) it is also easy to find the solution in the case of BTZ black holes. It is simply
\be
\phi \ = \ \frac{1}{\alpha}\arcsinh\left(\cos\chi\cot\theta\right)\,.
\ee

In the case of non-compact coordinates~(\ref{globalBTZ}) one needs to solve the following differential equation
\be
\cot\theta \left({\rm e}^{2 \rho }+J\right) \sqrt{\left({\rm e}^{2 \rho }-J\right)^2 {\phi'}^2+4 {\rm e}^{2 \rho }}+\left({\rm e}^{2 \rho }-J\right)^2 \phi' \ =\ 0\,.
\ee
It helps to know the solution in the compact coordinate. First we redefine the variable to get rid of $J$:
\be
\rho \to \rho +\frac12\,\log(- J)\,, \qquad \phi \to \frac{\phi}{\sqrt{-J}}\,.
\ee
The equation is now
\be
\cot\theta \left({\rm e}^{2 \rho }-1\right) \sqrt{\left({\rm e}^{2 \rho }+1\right)^2 {\phi'}^2+4 {\rm e}^{2 \rho }}+\left({\rm e}^{2 \rho }+1\right)^2 \phi' \ =\ 0\,.
\ee
This equation can be solved passing to the compact variables. In particular, one finds the solution
\be
\phi \ = \ \arcsinh\left(\frac{\cot\theta}{\cosh\rho}\right).
\ee
For the black hole of arbitrary mass one gets
\be
\phi \ = \ \frac1{\sqrt{-J}} \arcsinh\left(\frac{2\sqrt{-J}\cot\theta}{{\rm e}^\rho - J {\rm e}^{-\rho}}\right)
\ee
Examples of the profiles for three values of $\theta$ are shown on figure~\ref{fig:gads}.

\begin{figure}[htb]
 \centering
 \begin{minipage}{0.4\linewidth}
  \includegraphics[width=\linewidth]{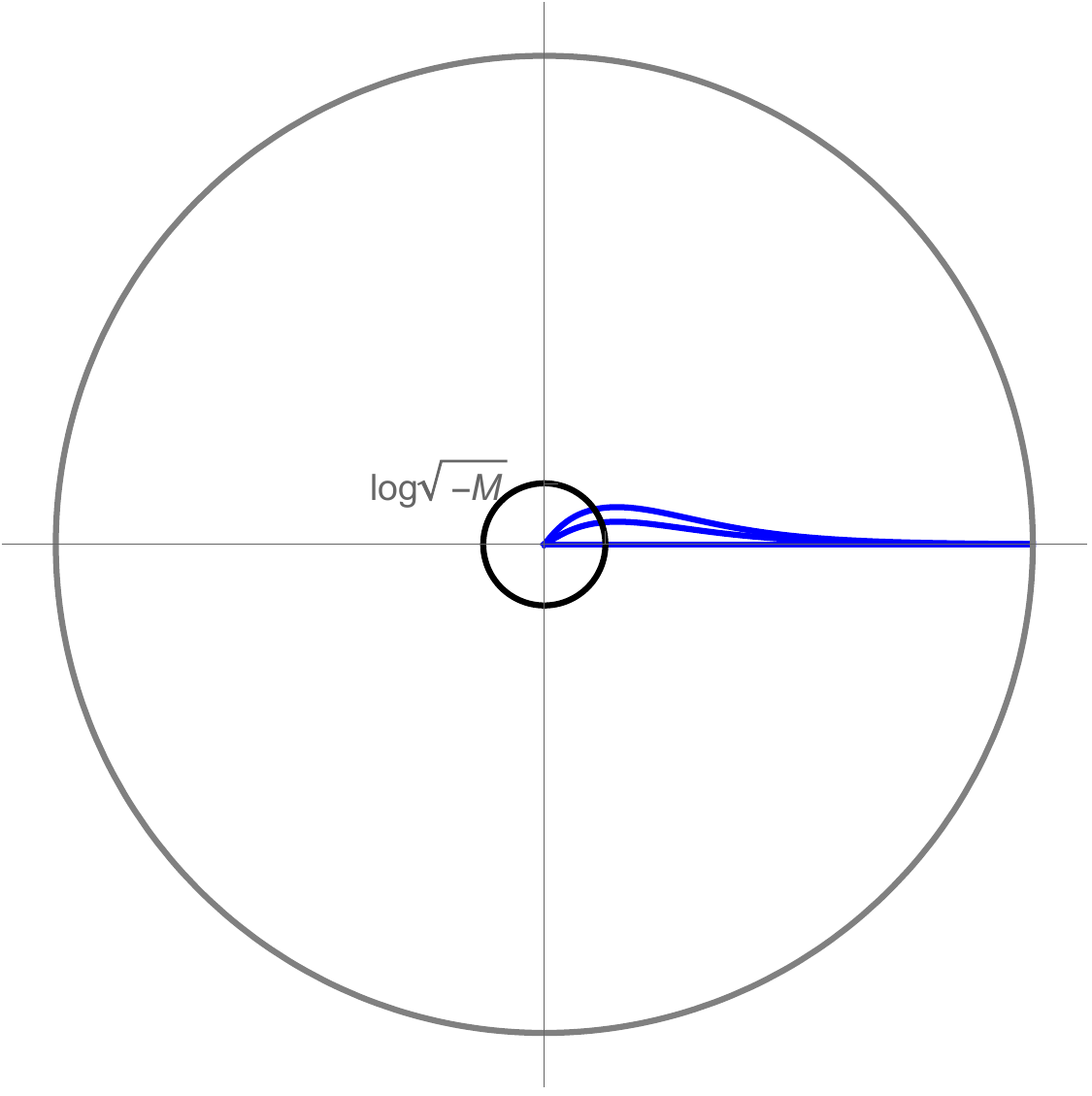}
 \end{minipage}
\hfill{
\begin{minipage}{0.4\linewidth}
  \includegraphics[width=\linewidth]{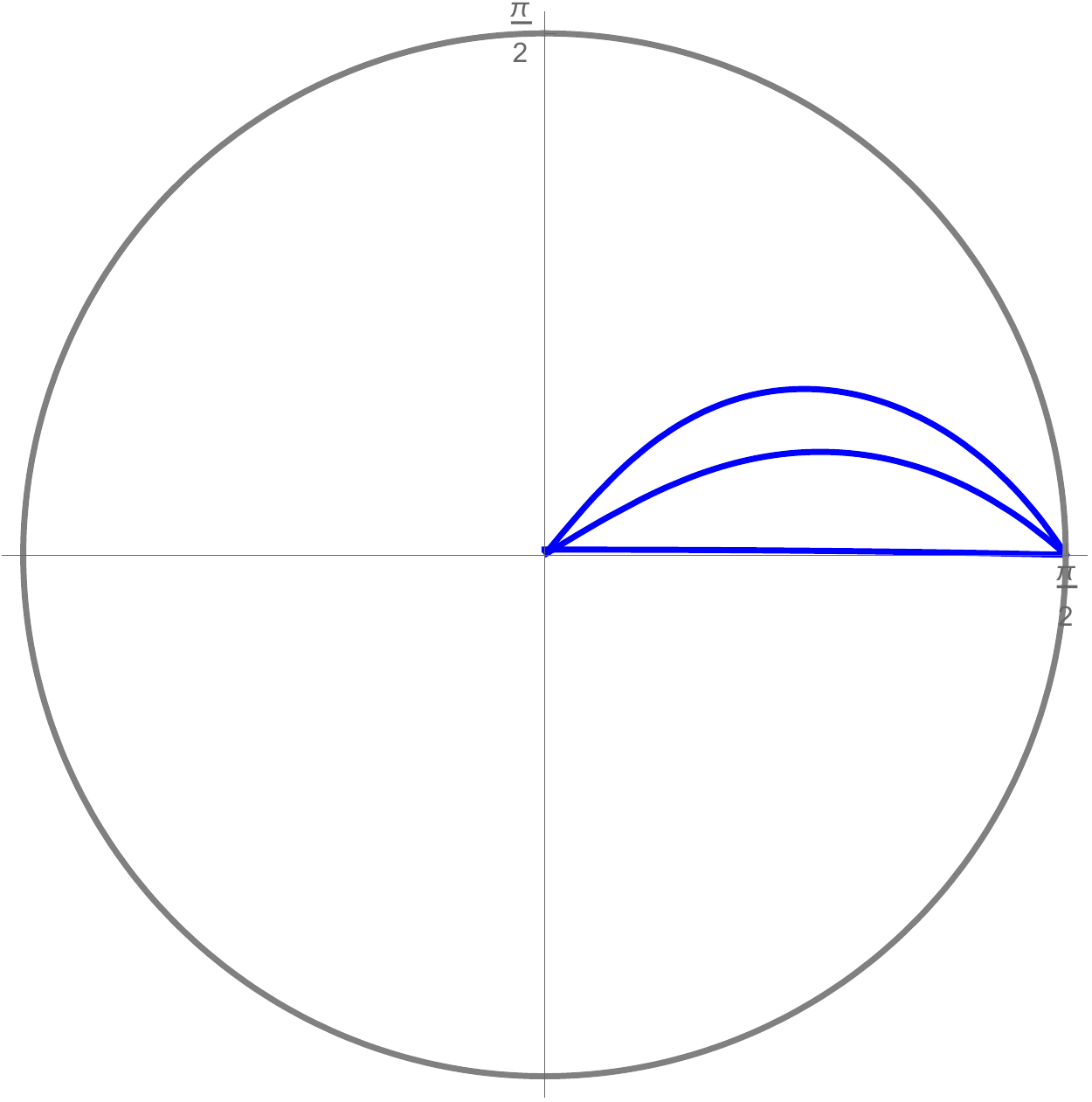}
 \end{minipage}
}
 \caption{Profiles of $Q$ in global BTZ black hole geometry in non-compact (left) and compact (right) coordinates for $-J=\alpha=1$ and $\theta=\pi/2,\pi/3,\pi/4$.}
 \label{fig:gbtz}
\end{figure}

\end{appendix}


\begin{thebibliography}{99}

\bibitem{AdS/CFT} J.~M.~Maldacena,
  Int.\ J.\ Theor.\ Phys.\  {\bf 38} (1999) 1113
   [Adv.\ Theor.\ Math.\ Phys.\  {\bf 2} (1998) 231]
  [hep-th/9711200];

  S.~S.~Gubser, I.~R.~Klebanov and A.~M.~Polyakov,
  Phys.\ Lett.\ B {\bf 428} (1998) 105
  [hep-th/9802109].

  E.~Witten,
  Adv.\ Theor.\ Math.\ Phys.\  {\bf 2} (1998) 253
  [hep-th/9802150].

\bibitem{Takayanagi:2011zk}
  T.~Takayanagi,
  Phys.\ Rev.\ Lett.\  {\bf 107} (2011) 101602
  [arXiv:1105.5165 [hep-th]].


\bibitem{Cardy:2004hm}
  J.~L.~Cardy,
  hep-th/0411189.

J.~L.~Cardy,
  Nucl.\ Phys.\ B {\bf 324} (1989) 581.

 \bibitem{AdS/BCFT2} M.~Fujita, T.~Takayanagi and E.~Tonni,
  JHEP {\bf 1111} (2011) 043
  [arXiv:1108.5152 [hep-th]].
  

\bibitem{HoloKondo}

  J.~Erdmenger, M.~Flory, C.~Hoyos, M.~N.~Newrzella and J.~M.~S.~Wu,
  Fortsch.\ Phys.\  {\bf 64} (2016) 109
  [arXiv:1511.03666 [hep-th]].

  J.~Erdmenger, M.~Flory, C.~Hoyos, M.~N.~Newrzella, A.~O'Bannon and J.~Wu,
  Fortsch.\ Phys.\  {\bf 64} (2016) 322
  [arXiv:1511.09362 [hep-th]].

\bibitem{Chu:2017aab}
  C.~S.~Chu, R.~X.~Miao and W.~Z.~Guo,
  JHEP {\bf 1704} (2017) 089
  [arXiv:1701.07202 [hep-th]].

\bibitem{AdSBCFTexamples} 

  R.~X.~Miao and C.~S.~Chu,
  JHEP {\bf 1803} (2018) 046
  [arXiv:1706.09652 [hep-th]].

  D.~Seminara, J.~Sisti and E.~Tonni,
  JHEP {\bf 1711} (2017) 076
  [arXiv:1708.05080 [hep-th]];

  D.~Seminara, J.~Sisti and E.~Tonni,
  JHEP {\bf 1808} (2018) 164
  [arXiv:1805.11551 [hep-th]].



  R.~X.~Miao,
  arXiv:1806.10777 [hep-th].

\bibitem{Freedan}
  I.~Affleck and A.~W.~W.~Ludwig,
  Phys.\ Rev.\ Lett.\  {\bf 67} (1991) 161.

D.~Friedan and A.~Konechny,
  Phys.\ Rev.\ Lett.\  {\bf 93} (2004) 030402
  [hep-th/0312197].

\bibitem{Universality}

  J.~L.~Cardy and I.~Peschel,
  Nucl.\ Phys.\ B {\bf 300} (1988) 377.

  C.~Holzhey, F.~Larsen and F.~Wilczek,
  Nucl.\ Phys.\ B {\bf 424} (1994) 443
  [hep-th/9403108].

  V.~E.~Korepin,
  Phys.\ Rev.\ Lett.\  {\bf 92} (2004) 096402.

\bibitem{RT}  S.~Ryu and T.~Takayanagi,
  Phys.\ Rev.\ Lett.\  {\bf 96} (2006) 181602
  [hep-th/0603001].

 \bibitem{Magan:2014dwa}
  J.~M.~Mag\'an, D.~Melnikov and M.~R.~O.~Silva,
  JHEP {\bf 1411} (2014) 069
  [arXiv:1408.2580 [hep-th]].

\bibitem{BulkImp} G.~C.~Levine,
  Phys.\ Rev.\ Lett.\ {\bf 93} (2004) 226402;

   I.~Peschel,
   J.\ Phys.\ A\ {\bf 38} (2005) 4327;

   J.~Zhao, I.~Peschel and X.~Q.~Wang,
    Phys.\ Rev.\ B {73} (2006) 024417.

\bibitem{Affleck:review} Ian Affleck, Nicolas Laflorencie, Erik S. Sorensen,
  J.\ Phys.\ A: Math.\ Theor.\ {\bf 42} (2009) 504009.
  [arXiv:0906.1809 [cond-mat.stat-mech]]



\bibitem{Erdmenger:2014xya}
  J.~Erdmenger, M.~Flory and M.~N.~Newrzella,
  JHEP {\bf 1501} (2015) 058
  [arXiv:1410.7811 [hep-th]].

\bibitem{holodefects} 

  A.~Karch and L.~Randall,
  JHEP {\bf 0106} (2001) 063
  [hep-th/0105132].

  O.~DeWolfe, D.~Z.~Freedman and H.~Ooguri,
  Phys.\ Rev.\ D {\bf 66} (2002) 025009
  [hep-th/0111135].

  D.~Bak, M.~Gutperle and S.~Hirano,
  JHEP {\bf 0305} (2003) 072
  [hep-th/0304129].

  E.~D'Hoker, J.~Estes and M.~Gutperle,
  JHEP {\bf 0706} (2007) 021
  [arXiv:0705.0022 [hep-th]].

  E.~D'Hoker, J.~Estes and M.~Gutperle,
  JHEP {\bf 0706} (2007) 022
  [arXiv:0705.0024 [hep-th]].

  O.~Aharony, L.~Berdichevsky, M.~Berkooz and I.~Shamir,
  Phys.\ Rev.\ D {\bf 84} (2011) 126003
  [arXiv:1106.1870 [hep-th]].

  M.~Chiodaroli, E.~D'Hoker and M.~Gutperle,
  JHEP {\bf 1202} (2012) 005
  [arXiv:1111.6912 [hep-th]].



\bibitem{QHE}
  M.~Fujita, M.~Kaminski and A.~Karch,
  JHEP {\bf 1207} (2012) 150
  [arXiv:1204.0012 [hep-th]].

  D.~Melnikov, E.~Orazi and P.~Sodano,
  JHEP {\bf 1305} (2013) 116
  [arXiv:1211.1416 [hep-th]].

\bibitem{TopDownExamples}
  M.~Chiodaroli, M.~Gutperle and L.~Y.~Hung,
  JHEP {\bf 1009} (2010) 082
  [arXiv:1005.4433 [hep-th]].

  M.~Chiodaroli, M.~Gutperle, L.~Y.~Hung and D.~Krym,
  Phys.\ Rev.\ D {\bf 83} (2011) 026003
  [arXiv:1010.2758 [hep-th]].

  M.~Chiodaroli, E.~D'Hoker and M.~Gutperle,
  JHEP {\bf 1207} (2012) 177
  [arXiv:1205.5303 [hep-th]].

  M.~Gutperle and J.~Samani,
  Phys.\ Rev.\ D {\bf 86} (2012) 106007
  [arXiv:1207.7325 [hep-th]].

  K.~Jensen and A.~O'Bannon,
  Phys.\ Rev.\ D {\bf 88} (2013) no.10,  106006
  [arXiv:1309.4523 [hep-th]].

  J.~Estes, K.~Jensen, A.~O'Bannon, E.~Tsatis and T.~Wrase,
  JHEP {\bf 1405} (2014) 084
  [arXiv:1403.6475 [hep-th]].

 \bibitem{Nagasaki:2011ue}
  K.~Nagasaki, H.~Tanida and S.~Yamaguchi,
  JHEP {\bf 1201} (2012) 139
  [arXiv:1109.1927 [hep-th]]. 

\bibitem{AdS/BCFT3}

  R.~X.~Miao, C.~S.~Chu and W.~Z.~Guo,
  Phys.\ Rev.\ D {\bf 96} (2017) no.4,  046005
  [arXiv:1701.04275 [hep-th]].

  A.~Faraji Astaneh and S.~N.~Solodukhin,
  Phys.\ Lett.\ B {\bf 769} (2017) 25
  [arXiv:1702.00566 [hep-th]].

  A.~Faraji Astaneh, C.~Berthiere, D.~Fursaev and S.~N.~Solodukhin,
  Phys.\ Rev.\ D {\bf 95} (2017) no.10,  106013
  [arXiv:1703.04186 [hep-th]].


\bibitem{Nozaki:2012qd}
  M.~Nozaki, T.~Takayanagi and T.~Ugajin,
  JHEP {\bf 1206} (2012) 066
  [arXiv:1205.1573 [hep-th]].


  
  
\bibitem{Israel} 
  W.~Israel,
  Nuovo Cim.\ B {\bf 44S10} (1966) 1
   [Nuovo Cim.\ B {\bf 44} (1966) 1]
   Erratum: [Nuovo Cim.\ B {\bf 48} (1967) 463];

  J.~P.~S.~Lemos and G.~M.~Quinta,
  Phys.\ Rev.\ D {\bf 89} (2014) no.8,  084051
  [arXiv:1403.0579 [gr-qc]].


\bibitem{CC}
  P.~Calabrese and J.~L.~Cardy,
  J.\ Stat.\ Mech.\  {\bf 0406} (2004) P06002
  [hep-th/0405152].


\bibitem{Azeyanagi:2007qj}
  T.~Azeyanagi, A.~Karch, T.~Takayanagi and E.~G.~Thompson,
  JHEP {\bf 0803} (2008) 054
  [arXiv:0712.1850 [hep-th]].




\bibitem{BTZ}
  M.~Banados, C.~Teitelboim and J.~Zanelli,
  Phys.\ Rev.\ Lett.\  {\bf 69} (1992) 1849
  [hep-th/9204099].

\bibitem{strip}
  S.~Ryu and T.~Takayanagi,
  JHEP {\bf 0608} (2006) 045
  [hep-th/0605073].

  T.~Nishioka, S.~Ryu and T.~Takayanagi,
  J.\ Phys.\ A {\bf 42} (2009) 504008
  [arXiv:0905.0932 [hep-th]].

\bibitem{Erdmenger:2017pfh}

  W.~Fischler and S.~Kundu,
  JHEP {\bf 1305} (2013) 098
  [arXiv:1212.2643 [hep-th]].

J.~Erdmenger and N.~Miekley,
  JHEP {\bf 1803} (2018) 034
  [arXiv:1709.07016 [hep-th]].

\bibitem{Headrick:2010zt}
  M.~Headrick,
  Phys.\ Rev.\ D {\bf 82} (2010) 126010
  [arXiv:1006.0047 [hep-th]].

\bibitem{Kitaev:2005dm}
  A.~Kitaev and J.~Preskill,
  Phys.\ Rev.\ Lett.\  {\bf 96} (2006) 110404
  [hep-th/0510092].

 \bibitem{Almheiri:2018ijj}
  A.~Almheiri, A.~Mousatov and M.~Shyani,
  arXiv:1803.04434 [hep-th].




\end{thebibliography}
\end{document}